\documentclass[aps,pra,reprint,superscriptaddress,amsmath,amsfonts,amssymb]{revtex4-1}

\usepackage{bm}
\usepackage[pdftex]{graphicx}
\usepackage{color}
\usepackage{bbold}
\usepackage{esint}
\usepackage{xargs}[2008/03/08]

\global\long\def\d{{\rm d}}
\newcommandx\tp[1][usedefault, addprefix=\global, 1=]{t_{#1}}
\newcommandx\rk[1][usedefault, addprefix=\global, 1=k]{r_{#1}}
\global\long\def\delt{{\rm d}t}
\newcommandx\m[1][usedefault, addprefix=\global, 1=k]{M(\rk[#1])}
\newcommandx\mdr[1][usedefault, addprefix=\global, 1=k]{M^{\dagger}(\rk[#1])}
\newcommandx\uk[1][usedefault, addprefix=\global, 1=\delt]{U(#1)}

\global\long\def\rt{R}
\newcommandx\var[1][usedefault, addprefix=\global, 1=]{\tau^{#1}}
\global\long\def\sz{\boldsymbol{\hat{Z}}}
\newcommandx\ot[1][usedefault, addprefix=\global, 1=]{\mathcal{M}_{\rt}^{#1}}
\newcommandx\rks[1][usedefault, addprefix=\global, 1=k]{r_{#1}^{2}}
\newcommandx\fr[1][usedefault, addprefix=\global, 1=k]{f(\rk[#1])}
\global\long\def\e{{\rm e}}
\global\long\def\etr{\mathcal{E}_{\rt}}
\newcommandx\ukd[1][usedefault, addprefix=\global, 1=\delt]{U^{\dagger}(#1)}
\newcommandx\cij[1][usedefault, addprefix=\global, 1=ij]{c_{#1}}
\newcommandx\sij[2][usedefault, addprefix=\global, 1=i, 2=j]{{\hat \sigma}_{#1}\otimes{\hat \sigma}_{#2}}
\global\long\def\id{I\otimes I}
\global\long\def\tr{{\rm Tr}}
\global\long\def\ri{\rho_{0}}
\newcommandx\zt[1][usedefault, addprefix=\global, 1=t]{\sz(#1)}
\newcommandx\aij[2][usedefault, addprefix=\global, 1=ij, 2=]{\alpha_{#1}(\tp[#2])}
\newcommandx\pc[1][usedefault, addprefix=\global, 1=ij]{\partial_{#1}}
\global\long\def\ip{i^{\prime}}
\global\long\def\jp{j^{\prime}}
\global\long\def\ijp{\ip\jp}
\newcommandx\fij[1][usedefault, addprefix=\global, 1={ij,\ijp}]{F_{#1}}
\global\long\def\dt{\d t}
\newcommandx\dtp[1][usedefault, addprefix=\global, 1=]{\d\tp[#1]}
\global\long\def\mt{\mathcal{T}}
\newcommandx\rtp[1][usedefault, addprefix=\global, 1=]{\rk[#1]}
\newcommandx\otr[1][usedefault, addprefix=\global, 1=]{\mathcal{M}_{R}^{#1}}
\global\long\def\sint{\int_{0}^{T}}
\newcommandx\ztp[1][usedefault, addprefix=\global, 1=]{\zt[{\tp[#1]}]}
\global\long\def\mint{\sint\cdots\sint}
\global\long\def\pr{P(\rt|\ri)}
\newcommandx\drtp[1][usedefault, addprefix=\global, 1=]{\d\rtp[#1]}
\newcommandx\frtp[1][usedefault, addprefix=\global, 1=]{f(\rtp[#1])}
\newcommandx\dett[2][usedefault, addprefix=\global, 1=, 2=]{\delta(\tp[#1]-\tp[#2])}
\global\long\def\avgo#1{\langle#1\rangle_{0}}
\global\long\def\ib{\boldsymbol{\hat{I}}}

\newcommandx\ccij[2][usedefault, addprefix=\global, 1=i, 2=j]{c_{#1#2}}

\newcommandx\ei[1][usedefault, addprefix=\global, 1=i]{\hat E_{#1}}

\global\long\def\rp{\rho^{\prime}}
\newcommandx\li[2][usedefault, addprefix=\global, 1=\rp, 2=X]{\mathcal{L}(#1|#2)}
\newcommandx\qp[1][usedefault, addprefix=\global, 1=\rp]{q(#1)}
\newcommandx\pp[1][usedefault, addprefix=\global, 1=\rp|X]{P(#1)}
\global\long\def\rha{\check{\rho}}
\newcommandx\cih[1][usedefault, addprefix=\global, 1=i]{\check{c}_{#1}}
\global\long\def\vci{\boldsymbol{c}}

\newcommandx\mbe[1][usedefault, addprefix=\global, 1=\cdot]{\mathbb{E}[#1]}
\global\long\def\mse{{\rm MSE}}
\global\long\def\cov{{\rm Cov}}
\global\long\def\vf{\boldsymbol{F}}
\global\long\def\vnb{\boldsymbol{\nabla}}
\newcommandx\pci[1][usedefault, addprefix=\global, 1=i]{\frac{\partial}{\partial c_{#1}}}

\global\long\def\fid{F(\rha,\rho)}
\global\long\def\trd{D(\rha,\rho)}

\global\long\def\vcb{\overline{\vci}}
\global\long\def\drp{\d\rp}

\global\long\def\vcp{\vci^{\prime}}

\newcommandx\oi[1][usedefault, addprefix=\global, 1=i]{o_{#1}}

\newcommandx\lij[1][usedefault, addprefix=\global, 1=ij]{\lambda_{#1}}
\newcommandx\nij[1][usedefault, addprefix=\global, 1=ij]{N_{#1}}
\global\long\def\vu{\overrightarrow{u}}
\global\long\def\vv{\overrightarrow{v}}

\newcommand{\X}{\hat X}
\newcommand{\Y}{\hat Y}
\newcommand{\Z}{\hat Z}
\newcommand{\I}{\hat I}
\newcommand{\IX}{{\hat I} \otimes {\hat X}}
\newcommand{\IY}{{\hat I} \otimes {\hat Y}}
\newcommand{\IZ}{{\hat I} \otimes {\hat Z}}
\newcommand{\XI}{{\hat X} \otimes {\hat I}}
\newcommand{\XX}{{\hat X} \otimes {\hat X}}

\newcommand{\YX}{{\hat Y} \otimes {\hat X}}
\newcommand{\YY}{{\hat Y} \otimes {\hat Y}}
\newcommand{\YZ}{{\hat Y} \otimes {\hat Z}}
\newcommand{\ZI}{{\hat Z} \otimes {\hat I}}

\newcommand{\ZY}{{\hat Z} \otimes {\hat Y}}
\newcommand{\ZZ}{{\hat Z} \otimes {\hat Z}}

\newcommand{\la}{\langle}
\newcommand{\ra}{\rangle}
\newcommand{\be}{\begin{equation}}
\newcommand{\ee}{\end{equation}}

\begin{document}

\title{Quantum state tomography with time-continuous  measurements: reconstruction with resource limitations}
\author{Areeya Chantasri}
\affiliation{Department of Physics and Astronomy, University of Rochester, Rochester, New York 14627, USA}
\affiliation{Center for Coherence and Quantum Optics, University of Rochester, Rochester, New York 14627, USA}
\affiliation{Centre for Quantum Computation and Communication Technology (Australian Research Council), Centre for Quantum Dynamics, Griffith University, Nathan, QLD 4111, Australia}
\author{Shengshi Pang}
\affiliation{Department of Physics and Astronomy, University of Rochester, Rochester, New York 14627, USA}
\affiliation{Center for Coherence and Quantum Optics, University of Rochester, Rochester, New York 14627, USA}
\author{Teerawat Chalermpusitarak}
\affiliation{Optical and Quantum Physics Laboratory, Department of Physics, Faculty of Science, Mahidol University, Bangkok 10400, Thailand}
\author{Andrew N. Jordan}
\affiliation{Department of Physics and Astronomy, University of Rochester, Rochester, New York 14627, USA}
\affiliation{Center for Coherence and Quantum Optics, University of Rochester, Rochester, New York 14627, USA}
\affiliation{Institute for Quantum Studies, Chapman University, 1 University Drive, Orange, California 92866, USA}
\date{\today}
\begin{abstract}
We propose and analyze quantum state estimation (tomography) using continuous quantum measurements with resource limitations, allowing the global state of many qubits to be constructed from only measuring a few. We give a proof-of-principle investigation demonstrating successful tomographic reconstruction of an arbitrary initial quantum state for three different situations:  single qubit, remote qubit, and two interacting qubits.  The tomographic reconstruction utilizes only a continuous weak probe of a single qubit observable, a fixed coupling Hamiltonian, together with single-qubit controls.  In the single qubit case, a combination of the continuous measurement of an observable and a Rabi oscillation is sufficient to find all three unknown qubit state components. For two interacting qubits, where only one observable of the first qubit is measured, the control Hamiltonian can be implemented to transfer all quantum information to the measured observable, via the qubit-qubit interaction and Rabi oscillation controls applied locally on each qubit. We discuss different sets of controls by analyzing the unitary dynamics and the Fisher information matrix of the estimation in the limit of weak measurement, and simulate tomographic results numerically. As a result, we obtained reconstructed state fidelities in excess of 0.98 with a few thousand measurement runs.
\end{abstract}

\maketitle

\section{Introduction}
Quantum state tomography is a general method for reconstructing a quantum state based on a collection of measurements made on a quantum system, all starting from the unknown state of interest \cite{Newton1968}. State tomography is one of the most important parts in building quantum information processing devices, as it is used in verifying quantum states before, during, or after any of the devices' processes \cite{Nielsen2000}. Standard procedures to implement tomography in the laboratory mainly aim to approximate textbooks projective measurements applied on separate ensembles, in which the measured observables are chosen carefully to assure informational completeness for the unknown quantum states \cite{Jamestomo2001,Doherty2008,Grosssensing2010,RenessicPOVM2016}. However, researchers have also explored different measurement and estimation approaches to achieve the same end, for example, using a sequential unsharp measurement \cite{Bassa2015}, or continuous weak measurement \cite{SilberfarbPRL2005,Smith2006,RiofrioJoPB2011,Six2016,Shojaee2018} in combination with estimation techniques such as Bayesian or maximum likelihood methods to estimate initial states of the measurement processes \cite{Blume-Kohout2010,Hradil2004}.


Of particular interest to this paper are investigations using continuous quantum measurements, of the type pioneered in recent experimental works \cite{Kater2013,Roch2014,ibarcq2015,Weberreview2016,Shay2016noncom,Naghiloo2016} and based on by now well-established theoretical methods such as the stochastic Schr\"odinger equation, stochastic master equation \cite{BookBachielli,BookCarmichael,wiseman2009quantum,BookJacobs}, and stochastic path integral \cite{Chantasri2013,chantasri2015stochastic}. We note that the weak continuous measurement is effectively a projective measurement after a long enough data-collection time \cite{wiseman2009quantum,Gross2015}, but the advantage of the former over the later is that the model is more practical in experiments and it allows us to include additional feedback controls to the systems, in order to achieve desired processes \cite{Wiseman1994,Doherty2000}.  The work by Six, \emph{et al.}~\cite{Six2016} theoretically and experimentally demonstrates quantum state tomography using continuous measurements and the maximum likelihood method on Rydberg atoms and superconducting circuit experiments. We highlight in particular the works of the group of I. Deutsch and P. S. Jessen, concerning tomographic reconstruction of initial collective state of an ensemble of atomic spins using continuous measurements \cite{RiofrioJoPB2011,SilberfarbPRL2005,Smith2006,Shojaee2018}. In these works, the measured observables are fixed and the information about other observables of interest is continually mapped onto the measured ones via continuous controlled unitary dynamics. We adapt this idea to measurements on individual qubits, using continuous measurement limited to only a single observable and continuous controls, in order to extract information about unknown initial states. Measurement backaction must be accounted for to accurately reconstruct the state. Closely related to the problem of quantum state tomography is that of parameter estimation, for which there have been a series of works dedicated to estimating Hamiltonian parameters from data acquired by a continuous measurement sequence \cite{whea10,Yonezawa1514,ralph2011frequency,Kiilerich2016,Luisparaest2017,Ralph2017}.

In this paper, we take up the problem of continuous measurement tomography applied to a network of qubits. Our aim is to show that global state reconstruction is possible using a fixed set of control parameters in a continuous way, even when all qubits are not measured. The main advantage of such a technique is that in contrast to projective measurement tomography, where a sequence of different gate operations is required to then read out one of many different observables (each of which is repeated many times),  in our continuous measurement construction, global information about the entire state is encoded in each run of the continuous measurement.   Therefore, the tradeoff is that there is a only a bit of information about all of the observables in each run of the experiment, and the protocol must be repeated many times to reduce the statistical uncertainties below a given level.  Nevertheless, the experimental procedure becomes much simpler:  it is the same every time in principle because the local controls are fixed, and only the state reconstruction algorithms become more complicated to numerically implement.  An important problem then becomes how to choose the local controls and estimation methods in order to accurately find the initial state. 

Several different methods are introduced to solve the above problems.  We discuss a method of using commutators between the Hamiltonian of the system's unitary dynamics and the different observables to be estimated in order to chose appropriate values of the local controls that allow all elements of interest to be accessed by the readout channel. We also present the calculation of the Fisher information matrix about the initial quantum state, from which we can optimize the controls to minimize the statistical uncertainties.  Given chosen sets of controls, continuous measurement records are then used to compute the Bayesian and maximum likelihood  estimators for the unknown states.  We construct a mapping operator for all observed records, proportional to an element of a positive-operator-valued measure (POVM), and use it to compute Bayesian probability for all possible candidate states. This is to dispense with computing separate quantum trajectories for all candidate states which greatly speeds up the numerical estimation procedure. 

We illustrate our methods first in detail for the case of a single qubit, showing how the estimation procedures can be applied using quantum trajectory theory, and giving the tomography results given a local control and single component pseudo-spin readout.  We then incorporate a resource limitation, and show the same methods are able to reconstruct the state of a remote qubit that is coupled to the one being read out.  Finally, we demonstrate arbitrary full two qubit state reconstruction using local controls and fixed coupling.  We show that by choosing the local controls within a relatively broad range of parameters, good state fidelity with the target state can be easily reached.  

The paper is organized as follows.  In Sec.~\ref{sec-background}, we layout the background information for the continuous measurement of a single qubit's observable, and estimation techniques used for quantum state tomography. Our main results are presented in Sec.~\ref{sec-mainresults}, including the methods to choose local qubit controls in Sec.~\ref{sec-mainresults-method}, and the analysis of the three examples: single-qubit state tomography in Sec.~\ref{sec-mainresults-sqb}, remote-qubit state tomography in Sec.~\ref{sec-mainresults-tqb}, and two-qubit tomography in Sec.~\ref{sec-mainresults-tqb}, with results from numerical simulations. We conclude our results in Sec.~\ref{sec-conclusion} and present detailed calculation of the Fisher information matrix and numerical methods in the Appendices.

\section{Background}\label{sec-background}
\subsection{Time-continuous measurement of a qubit's observable}\label{sec-background-contmeas}
The concept of quantum measurement generalized to arbitrary strength and occurring over a finite time \cite{Diosi1988,BookBachielli} is becoming more recognized in the quantum information research community \cite{wiseman2009quantum,BookJacobs}. Not only that it describes more accurately the measurement processes in practical experiments, but also it opens possibilities for adding system controls during the measurement processes, such as feedback, in order to achieve desired outcomes more efficiently. Generalized quantum measurement theory considers a system of interest interacting with its environment (or detectors) where all or parts of the environment are observed. The evolution of the system depends on the system-environment interactions, as well as on the observed results of the measurement performed on the environment. 

In a limit where there is no information about the actual state of the environment, i.e., the environment is not observed or is averaged over, under the Markov assumption, the system's evolution is described by a Lindblad master equation \cite{Lind1976},
\begin{align}\label{eq-mastereq}
\partial_t \rho = -i [{\hat H} , \rho ] + \sum_k {\cal D}[{\hat c}_k]\rho,
\end{align} 
where $\rho$ is a density matrix representing a state of the system and ${\hat H}$ is a Hamiltonian describing unitary dynamics of the system. The Lindblad operator, defined as ${\cal D}[{\hat c}] \rho = {\hat c} \rho {\hat c}^{\dagger} - \frac{1}{2}({\hat c}^{\dagger} {\hat c} \rho + \rho {\hat c}^{\dagger} {\hat c} )$, describes the decoherence of the system's state as a result of the interaction with the environment, via different channels denoted by the Lindblad operators ${\hat c}_k$. 

In general, there can be many dephasing channels to consider; however, in this work we are interested in the interaction that leads to a measurement of an observable $\Z$ of a qubit, which might be part of a network of many qubits. Therefore, the system's dynamics only includes unitary dynamics (the first term in Eq.~\eqref{eq-mastereq}) and measurement backaction conditioning on particular measurement records. The backaction from continuous weak measurement of a single qubit has been studied quite extensively, largely motivated by experiments in quantum optics, and in solid state physics such as electronic states in double quantum dots. More recently, superconducting qubits coupled dispersively to microwave cavities have been investigated \cite{Kater2013,Roch2014}. 

We consider a single measurement record from an experiment. Detectors will typically return a discretized signal with a time step $\dt$ to get $R \equiv \{ r_1, r_2, ..., r_n\}$, where $r_k$ is an average signal from $t_{k-1}$ to $t_k = t_{k-1} + \dt$ and $n \,\dt = T$, assuming that $\dt$ is long enough for the Markov assumption to be valid, but short enough so that all types of system's dynamics commute with each other to first order of $\dt$. The dynamics for the system state under the measurement can be written in a Kraus form, given an initial state $\rho_0$,
\begin{equation}\label{eq-stateupdate}
\rho(t) = \frac{{\cal M}_R \rho_0 {\cal M}_{R}^{\dagger}}{{\rm Tr}[{\cal M}_R^{\dagger}{\cal M}_R \rho_0 ]},
\end{equation}
where we have used an operator ${\cal M}_R$ mapping the state from an initial time $t=0$ to any time $t$. The operator is defined as ${\cal M}_R  \equiv U(\dt) M(r_n)\cdots U(\dt)M(r_1)$, where $U(\dt) = \exp(- i {\hat H} \dt)$, describing both the unitary dynamics and measurement backaction from observing the record $R$. The operator $\m$ is a measurement operator, which, for our studies, can be calculated using the quantum Bayesian method introduced in \cite{Korotkov1999,Korotkov2001},
\begin{equation}\label{eq-measop}
\m=\text{\ensuremath{\Big(\frac{\delt}{2\pi\tau}\Big)}}^{\frac{1}{4}}\exp\left[-\frac{(\rk- {\bm \Z} )^{2}\delt}{4\tau}\right],
\end{equation}
describing an approximate Gaussian distribution of the measurement results. The observable ${\bm \Z} \equiv \Z_1 \otimes \I_2 \otimes \cdots \otimes \I_m$ represents the $z$ Pauli operator on the first qubit of the system of $m$ qubits, and can be considered as the dephasing channel ${\hat c}_k \propto {\hat {\bm Z}}$ in Eq.~\eqref{eq-mastereq}. The measurement strength is characterized by a measurement rate $\Gamma \equiv 1/2 \tau$, where $\tau$ in Eq.~\eqref{eq-measop} stands for a characteristic measurement time, i.e., a signal-integration time to reach unit signal-to-noise ratio \cite{Korotkov2001}. Therefore, given a measurement record $R$ and a known Hamiltonian ${\hat H}$, one can calculate a quantum trajectory $\rho(t)$ from an initial state $\rho_0$.

Since the measurement results are probabilistic, we can compute a probability density function for a record $R$ to occur given the initial state $\rho_0$,
\begin{align}\label{eq-probrecord}
P(R|\rho_0) = {\rm Tr}[{\cal M}_R^{\dagger} {\cal M}_R \rho_0 ],
\end{align} 
which is the same as the denominator of Eq.~\eqref{eq-stateupdate}. In the case that the initial state $\rho_0$ is unknown, a Bayesian probability density function of a possible initial state can be obtained as: $P(\rho' | R) \propto P(R|\rho') P(\rho')$ given a prior probability function of the unknown state $P(\rho')$. This Bayesian probability function for an unknown initial state is the main quantity used in the state estimation process.

We note that Eq.~\eqref{eq-stateupdate} is equivalent to the stochastic master equation \cite{wiseman2009quantum,BookJacobs} in the limit of $\dt \rightarrow 0$. Decoherence, inefficiencies, or extra dephasing effects can be added to the above model, which only results in degrading the quality of the state estimation.


\subsection{Introduction to Quantum state tomography}\label{sec-background-tomo}

In this section, we review some main approaches to quantum state tomography,
and briefly discuss about how to quantify the errors of tomography.
We will mainly focus on strategies of the linear inversion (LI), the maximum
likelihood estimation (MLE) and the Bayesian mean estimation (BME).

\subsubsection{Tomography strategies}

Quantum states can be characterized by a set of parameters which quantify the weights of different basis operators in the density matrix. The purpose of quantum state tomography is to find these parameters that determine the quantum
states. Suppose we want to reconstruct a quantum state in a $d$ dimensional
Hilbert space, and the density matrix of the state is $\rho$. We
can expand $\rho$ in an orthogonal matrix basis $\{\ei[0],\cdots,\ei[d^{2}-1]\}$
in the complex-matrix $\mathbb{C}^{d}\times\mathbb{C}^{d}$ space,
\begin{equation}
\rho=\frac{1}{d}\sum_{i=0}^{d^{2}-1}\cij[i]\ei[i],\label{eq:den mat para}
\end{equation}
where $\ei$'s satisfy $\tr(\ei\ei[j])=\delta_{ij}\I$ and $\ei^{\dagger}=\ei$.
It is straightforward to verify that
\begin{equation}
\cij[i]=\tr(\rho\,\ei),
\end{equation}
is a component of the state $\rho$ corresponding to an observable $\ei$.

Using a linear inversion method \cite{Newton1968}, by measuring each observable $\ei$, we can obtain an estimate of $\cij[i]$ from,
\begin{equation}
{\check c}_{i,{\rm LI}}=\frac{\sum_j\nij\lij}{\sum_j\nij},
\end{equation}
where $\lij$ are the eigenvalues of $\ei$, and $\nij$ are the number of times that the measurement result turns out to be $\lij$ when measuring $\ei$. We note that we use a variable with an upside-down hat ``$\check{a}$'' to represent an estimator of that variable  $a$. Due to the constraint, $\tr(\rho)=1$, we only need to measure $d^{2}-1$ of $\ei$'s. However, while this linear inversion is simple and efficient, a major defect of this approach is that the estimate of the
density matrix it generates is not always a legitimate one for quantum
systems. This is because the unknown coefficients $\cij[i]$'s are estimated separately
and the fluctuation in the estimates of the coefficients may make
the estimate of the density matrix non-physical. In order to resolve
this issue, we need to find ways to estimate the density matrix
of a quantum system within the positivity constraint, instead of estimating the coefficients
of the density matrix individually.

The maximum likelihood estimation (MLE) and the Bayesian mean estimation
(BME) can resolve the positivity issue. The key idea of these two approaches
is to assign prior probabilities to different possible density
matrices of the quantum system, so that invalid candidate states can be ruled out. 

Let us suppose that we have measurement data denoted by $X$, which can include
many measurement records obtained from measuring an ensemble of unknown quantum systems, and a likelihood function of an unknown quantum state $\rho'$ given the observed record $X$ is denoted by ${\cal L}(\rho'|X)$. The MLE \cite{Hradil2004} is to choose the density matrix which maximizes
the likelihood function as the estimate of the unknown state. Mathematically,
the estimate of the density matrix by MLE is given by,
\begin{equation}\label{eq-MLE}
{\check \rho}_{{\rm MLE}}=\arg\max_{\rp}\li,
\end{equation}
noting that the prior distribution for MLE is implicitly the uniform distribution over all legitimate candidate states where the likelihood function is maximized.

In the case if prior probability distribution $\qp$ is non-uniform over
the valid state space of $\rp$, we can compute the most probable Bayesian
estimator (MPBE) for the tomography,
\begin{equation}
{\check \rho}_{{\rm MPBE}}=\arg\max_{\rp}\qp\li,
\end{equation}
which differs from the MLE in explicitly taking a non-uniform prior distribution $\qp$ into account.

The benefit of MLE is that, if a large amount
of data is given, the MLE can asymptotically saturate the Cram\'er-Rao bound for parameter
estimation (which will be introduced in the next subsection). Therefore, in principle, the MLE method can reconstruct the unknown density matrix
with an optimal precision. However, the MLE also suffers from some
drawbacks. For example, the estimate may be rank
deficient, sensitive to initial choices of states used in maximization algorithms, and easily
trapped in local minimas.

On the other hand, the BME \cite{Blume-Kohout2010} approach, reconstructing the density matrix by averaging over all possible states
weighted with posterior probabilities, is capable to overcome the defects of the MLE method. Given that the prior probability
distribution for a candidate state $\rp$ is $\qp$, the posterior
distribution is,
\begin{equation}\label{eq-Bayesprob}
\pp=\frac{\qp\li}{\int\qp\li\drp},
\end{equation}
and the BME is then defined as
\begin{equation}\label{eq-BME}
{\check \rho}_{{\rm BME}}=\int\pp\rp\drp.
\end{equation}
We note that the advantage of BME is that it can provide full-rank estimate
of the density matrix as well as simple ways to calculate error bars of the estimate (see next section). Moreover, one can optimize the
accuracy of the state tomography measured by some operational divergence \cite{Blume-Kohout2010}.
However, this BME method requires sampling the whole space of the candidate states rather than only near the maximum as in Eq.~\eqref{eq-MLE}, which is generally a computationally intensive task.

\subsubsection{Benchmarking the performance of tomography}

To characterize the performance of a quantum state tomography scheme, we need to define error measures that properly quantifies any deviations of the estimated states from their corresponding true states. From the point of view of parameter estimation, quantum state tomography is an estimation problem, where the unknown parameters to be estimated are the coefficients of quantum states of interest. Therefore, we consider three types of error measures: the mean square error for the individual components, quantum state fidelity, and the trace distance.

The density matrix of a quantum system is characterized by parameters
$\cij[i]$'s, in Eq.~\eqref{eq:den mat para}. We denote their estimates by $\cih$'s and write the true and estimated parameters
in vector forms as, $\vci=\{\cij[0],\cdots,\cij[d^{2}-1]\}$ and ${\check{\bm c}}=\{\cih[0],\cdots,\cih[d^{2}-1]\}$.
The mean square errors of the estimators $\cih$'s can be described by an error
covariance matrix,
\begin{equation}
\mse=\mbe[({\check{\bm c}}-\vci)^{T}({\check{\bm c}}-\vci)],
\end{equation}
where $\mbe[\cdots]$ represents an average over all possible tomography results. If the estimation is unbiased, then $\mbe[{\check{\bm c}}]=\vci$ and
$\mse$ becomes a covariance matrix of ${\check{\bm c}}$,
\begin{equation}\label{eq-covmat}
\cov=\mbe[({\check{\bm c}}-{{\mbe[{\check{\bm c}}]}})^{T}({\check{\bm c}}-{{\mbe[{\check{\bm c}}]}})].
\end{equation}
In estimation theory, the covariance matrix of unbiased estimators of deterministic parameters has a lower bound set by the Cram\'er-Rao
bound \cite{Cramer1946},
\begin{equation}
\cov\geq\frac{1}{N}\vf^{-1},
\end{equation}
where $N$ is a number of repeated independent measurements, and $\vf$ is the Fisher information
matrix. Following the previous section, let us use $X$ to denote the measurement result where its statistics are described by a likelihood function ${\cal L}(X|\rho')$. The Fisher information matrix is defined as
\begin{equation}\label{eq-Fishermat}
\vf=\int\frac{\vnb^{T}{\cal L}(X|\rho')\vnb {\cal L}(X|\rho')}{{\cal L}(X|\rho')}\d X,
\end{equation}
where $\vnb$ is a vector of partial derivative operators
with respect to the parameters $\vci$, i.e., $\vnb=\{\pci[0],\cdots,\pci[d^{2}-1]\}$.

For the Bayesian mean estimation Eq.~\eqref{eq-BME}, one
can also define a covariance matrix based on the posterior probability as a measure of the uncertainty of the BME \cite{Blume-Kohout2010}. The Bayesian covariance matrix is given by,
\begin{equation}\label{eq-baycovmat}
{\rm Cov}_{\rm BME}=\int\pp(\vcp-\vcb)^{T}(\vcp-\vcb)\drp,
\end{equation}
where $\vcp$ is a component vector of the candidate state $\rp$, 
and $\vcb$ is a Bayesian mean defined as,
\begin{equation}
\vcb=\int\pp\vcp\drp.
\end{equation}
It is straightforward to verify that a component of this mean vector $\vcb$ is related to the Bayesian mean estimate of the quantum state Eq.~\eqref{eq-BME} via the relation: $ {\bar c}_i =\tr(\rha_{{\rm BME}}\ei)$. 


The covariance matrix defined in Eq.~\eqref{eq-covmat} characterizes the estimation errors of individual parameters and the correlation between them; however, it cannot give an overall benchmark for the quality of the quantum state reconstruction. We therefore need to use quantities that can measure quantum state distances to characterize
the overall error of quantum state tomography. Two typical measures
for the distance between two quantum states are the fidelity and the trace
distance \cite{Nielsen2000}. 

The quantum state fidelity between an estimate ${\check \rho}$ and
a true density matrix $\rho$ is defined as,
\begin{equation}\label{eq-fidelity}
F({\check \rho},\rho)=\tr\sqrt{\sqrt{\rho}{\check \rho}\sqrt{\rho}},
\end{equation}
whereas, the trace distance between ${\check \rho}$ and $\rho$ is
\begin{equation}
D({\check \rho}, \rho)=\frac{1}{2}\tr|{\check \rho}-\rho|.
\end{equation}
For a single qubit state, the trace distance is proportional to the Euclidean distance between the Bloch vectors of the two states. But for a general case, an important relation between the fidelity and the trace distance is,
\begin{equation}
1-F({\check \rho},\rho)\leq D({\check \rho},\rho) \leq\sqrt{1-F({\check \rho}, \rho)^{2}},
\end{equation}
which implies that, if $\fid$ is close to unity, $\trd$ is close to zero, and vice versa. The two measures are consistent in characterizing any discrepancies between two neighboring quantum states.

It should also be noted that the above error measures are dependent on
the actual true states of the quantum system. Therefore, to get a state-independent benchmark for the state tomography, the errors should be averaged over
all possible true states. Moreover, it is worth noting the difference between the Bayesian error
and the statistical error (which the Fisher information, the fidelity
and the trace distance rely on): the Bayesian error measures the uncertainty
between different candidate states in the Bayesian posterior distribution,
while the statistical error measures the uncertainty between different
estimates that arises from the fluctuation in different sets of measurement results.


\section{Quantum state tomography with continuous measurement}\label{sec-mainresults}

\begin{figure}
\includegraphics[width=8.8cm]{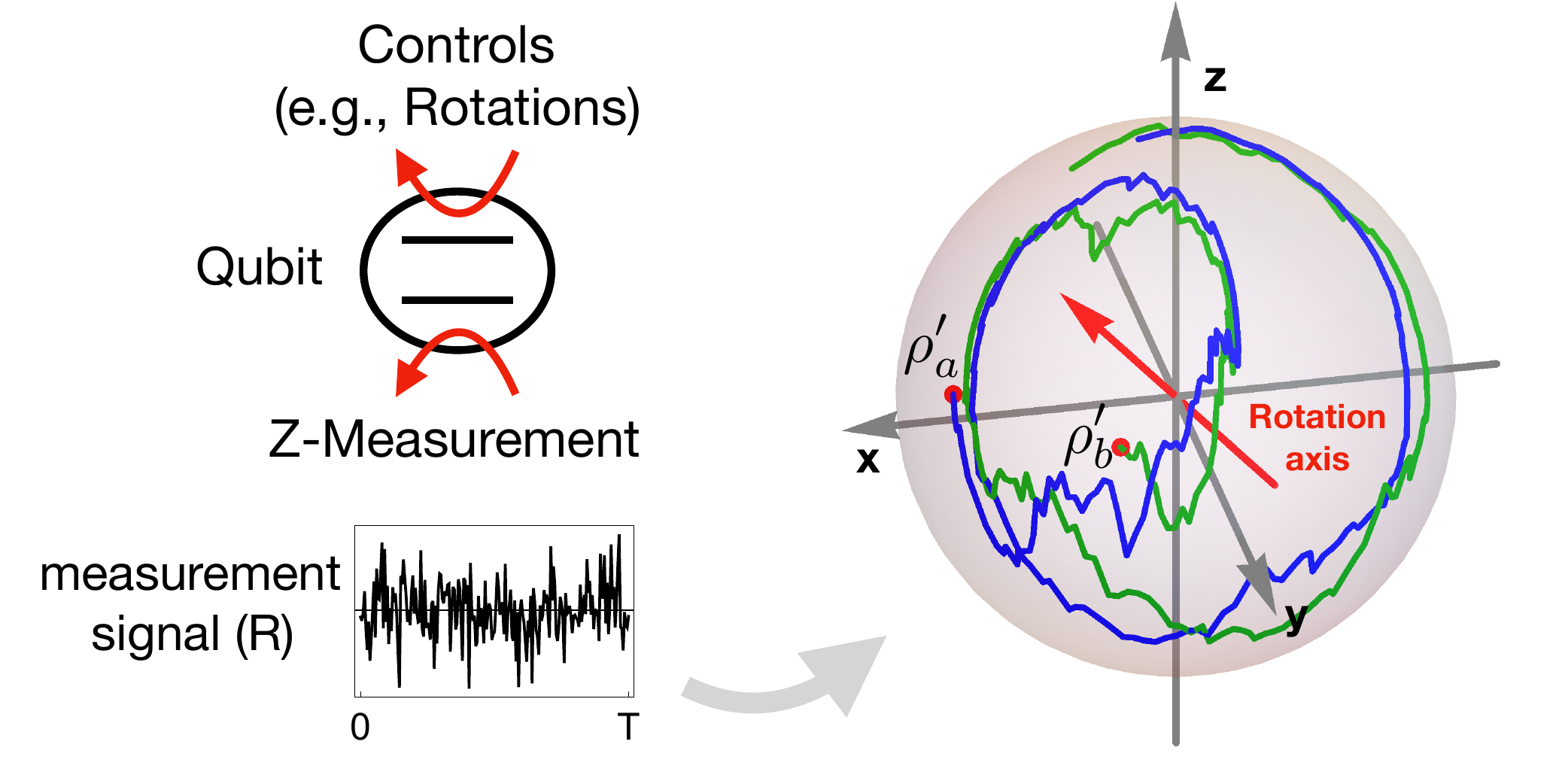}
\caption{Schematic diagram showing a single qubit tomography and trial qubit trajectories on a Bloch sphere. Left: A two-level system (qubit) under a continuous measurement of the $\Z$ observable and a controlled rotation. The continuous measurement for a duration of time $T$ gives a fluctuating record $R$, which is used in estimating the qubit's unknown initial state. Right: A Bloch sphere with two qubit trajectories (blue and green), generated using the same record from the left, but with different guessed initial states (shown as two red dots on the Bloch sphere). The two trajectories give two different values of the probability density (likelihood) function, $P(R|\rho'_{a})$ and $P(R|\rho'_{b})$.}
\label{fig-sqbDiagram}
\end{figure}

In the conventional linear inversion tomography, an ensemble of $N$ unknown systems prepared in the same state is divided into $p$ sets where projective measurements of $p$ observables are applied to the $p$ sets. However, we can perform measurements that can collect information about all observables first, and then use the methods presented in the previous section to compute the best estimator of the unknown state. Here, we make use of the continuous weak (finite strength) measurement of only one system's observable, in combination with continuous controls, to extract the information of all observables needed (see Figure~\ref{fig-sqbDiagram}).

With our continuous tomography strategy, we aim at performing the same experiment protocol to all $N$ systems, giving a set of measurement records $\{ R_j \} = \{ R_1, ..., R_N\}$. Since the $N$ records are independent of each other, the likelihood function for the measurement results is simply a product of Eq.~\eqref{eq-probrecord} for all records $R_j$,
\begin{align}\label{eq-likelihoodR}
{\cal L}(\rho'|\{R_j\})= \prod_j^N {\rm Tr}[{\cal M}_{R_j} \rho' {\cal M}_{R_j}^{\dagger}],
\end{align}
where we have used $\rho'$ as a candidate for the unknown initial state. If the measurement records contain information about parameters we need to estimate, i.e., the records can be written as a function of the parameters plus noninformative noises; then we can use the likelihood function Eq.~\eqref{eq-likelihoodR} or its corresponding Bayesian probability to compute estimators for the parameters using the technique from previous section.

The problem then becomes: how to choose the controls so that the measurement results are informationally complete for the tomography of the state. In the following subsections, we will elaborate on two techniques used in determining the controls, and present our investigation using three examples: a single qubit tomography, a remote qubit tomography, and a two qubit tomography, all depending only on the measurement of an observable $\Z$ of a qubit, and Rabi oscillations applied locally to each of the qubits, with fixed angles and rotation frequencies.


\subsection{Proposed methods to determine qubit controls}\label{sec-mainresults-method}
We present two approaches in this subsection to find suitable values of the controls to apply: a more crude but simpler approach using commutation relationships among the qubits' observables, and an approach using the Fisher information matrix of the unknown state parameters. Since we require the unitary controls to map all observables to the measured observable at some point during the measurement time, the unitary dynamics should dominate the measurement backaction. Considering only the unitary dynamics, for an infinitesimal time $\dt$, the system evolves as,
\begin{align}\label{eq-Hamevolve}
\nonumber \rho({\rm d}t)=&\, e^{-i {\hat H} \dt} \rho_0 e^{+i {\hat H} \dt} \\
\nonumber =&\, \frac{1}{d}\sum_j c_j^0 {\hat E}_j - i [{\hat H} , \frac{1}{d} \sum_j c_j^0 {\hat E}_j] \dt \\
&- \frac{\dt^2}{2}[{\hat H}, [ {\hat H}, \frac{1}{d} \sum_j c_j^0 {\hat E}_j]] + \cdots,
\end{align}
where we have substituted the state expansion defined in Eq.~\eqref{eq:den mat para} for the initial state $\rho_0$. The $\cdots$ refers to terms with higher orders of $\dt$ which contain three or more orders of commutators between the initial state $\rho_0$ and the system's Hamiltonian ${\hat H}$. The evolved quantum state $\rho(\dt)$ is shown as a function of the components $c_j^0$ of the initial state and is then measured in $\Z$ basis. 

If we also expand the left-hand side of Eq.~\eqref{eq-Hamevolve}, $\rho(\dt) = (1/d)\sum_k c_k^{\dt} {\hat E}_k$, in the same basis, we find a relationship
\begin{align}\label{eq-cZtransfer}
c_{{\bm \Z}}^{\dt} = c_{\bm \Z}^0 + \dt \sum_{j'} \alpha_{j'} c_{j'}^0  + \dt^2 \sum_{j''} \alpha_{j''} c_{j''}^0 + \cdots,
\end{align}
where  $c_{\bm \Z}^{\dt} = {\rm Tr}[\rho(\dt) {\bm \Z}]$ is the component of the system state's on the measured observable. The sums over indices $j'$ and $j''$ are for components $c_{j'}^0$ and $c_{j''}^0$ that satisfy $[{\hat H}, {\hat E}_{j'}]  \propto {\bm \Z}$ and $[{\hat H}, [{\hat H}, {\hat E}_{j''}]] \propto {\bm \Z}$, respectively, noting that any non-zero proportional constants are included in the definitions of $\alpha_{j'}$ and $\alpha_{j''}$. Therefore, the value of $c_{\bm \Z}^{\dt}$ is what is being measured, which we would like it to be a function of all components $c_j^0$ for $j = 1,2,..., d^2-1$ we need to estimate.


Another approach is to calculate the Fisher information matrix Eq.~\eqref{eq-Fishermat} for the parameters defining the unknown state. The probability distribution of a measurement result $R$ is given by Eq.~\eqref{eq-probrecord} and the integral $\int {\rm d}X$ in \eqref{eq-Fishermat} is then replaced with $\int {\rm d}r_1 \cdots {\rm d}r_n$. However, because the Kraus operator ${\cal M}_R$ is a product of noncommuting matrices $U(\dt)$ and $M(r_k)$ for $k = 1,...,n$, the product is not easily simplified. To obtain analytic results, we expand the measurement operator Eq.~\eqref{eq-measop} for small $\dt$, 
\begin{equation}
\m\approx \left( \frac{\dt}{2\pi \tau}\right)^{1/4}\e^{-\frac{(2 r_k^2+1)\dt}{4\var}}\Big({\bm \I}+\frac{\rk\delt}{2\var}{\bm \Z}\Big),\label{eq:appr meas op}
\end{equation}
where ${\bm \I} \equiv \I_1 \otimes \I_2 \cdots \otimes \I_m$ is an identity matrix for a system of $m$ qubits, and substitute the expansion to compute an approximated function of the probability distribution function $P(R|\rho_0)$ in Eq.~\eqref{eq-probrecord}. The $(j,j')$-th element of the Fisher information matrix is given by,
\begin{equation}\label{eq-Fishermat2}
F_{j,j'}=\dotsint\frac{\partial_j P(R|\rho_0)\partial_{j'} P(R|\rho_0)}{P(R|\rho_0)} {\rm d}r_0 \cdots {\rm d}r_n,
\end{equation}
where $j$ (and $j'$) are indices for different components $c_j^0$ (and $c_{j'}^0$) of the system state $\rho_0$ and the partial derivatives are defined as $\partial_j \equiv \partial/\partial c_j^0$. We show in Appendix~\ref{sec-fisher} the full derivation of the matrix for the two-qubit case. The analytical results we have are expansions to first order in $1/\tau$. Therefore, the results are applicable to weak measurement limit $\tau \gg T$. The calculation can also be generalized to the single qubit tomography and more. We will use these analyses to help us choose dynamical controls for the three examples in the following subsections.

\subsection{Single-qubit state tomography}\label{sec-mainresults-sqb}

The first application of our approach is quantum state tomography for an unknown single qubit state, where only the $\Z$ observable of the qubit can be accessed or be measured. Since we are interested in the qubit control being a simple Rabi oscillation along a fixed axis with constant frequency, there are only two degrees of freedom to vary: the axis of rotation and the rotation rate. The Hamiltonian of the system is given by,
\begin{align}
{\hat H} = \frac{\Omega}{2} ( {\vec n} \cdot {\hat {\vec \sigma}} ),
\end{align}
where $\Omega$ is the oscillation rate, $\vec n$ is a unit vector describing the axis of rotation, and ${\hat {\vec \sigma}} = \{ \X, \Y, \Z \}$ is a vector of Pauli operators.

\begin{figure}
\includegraphics[width=8.8cm]{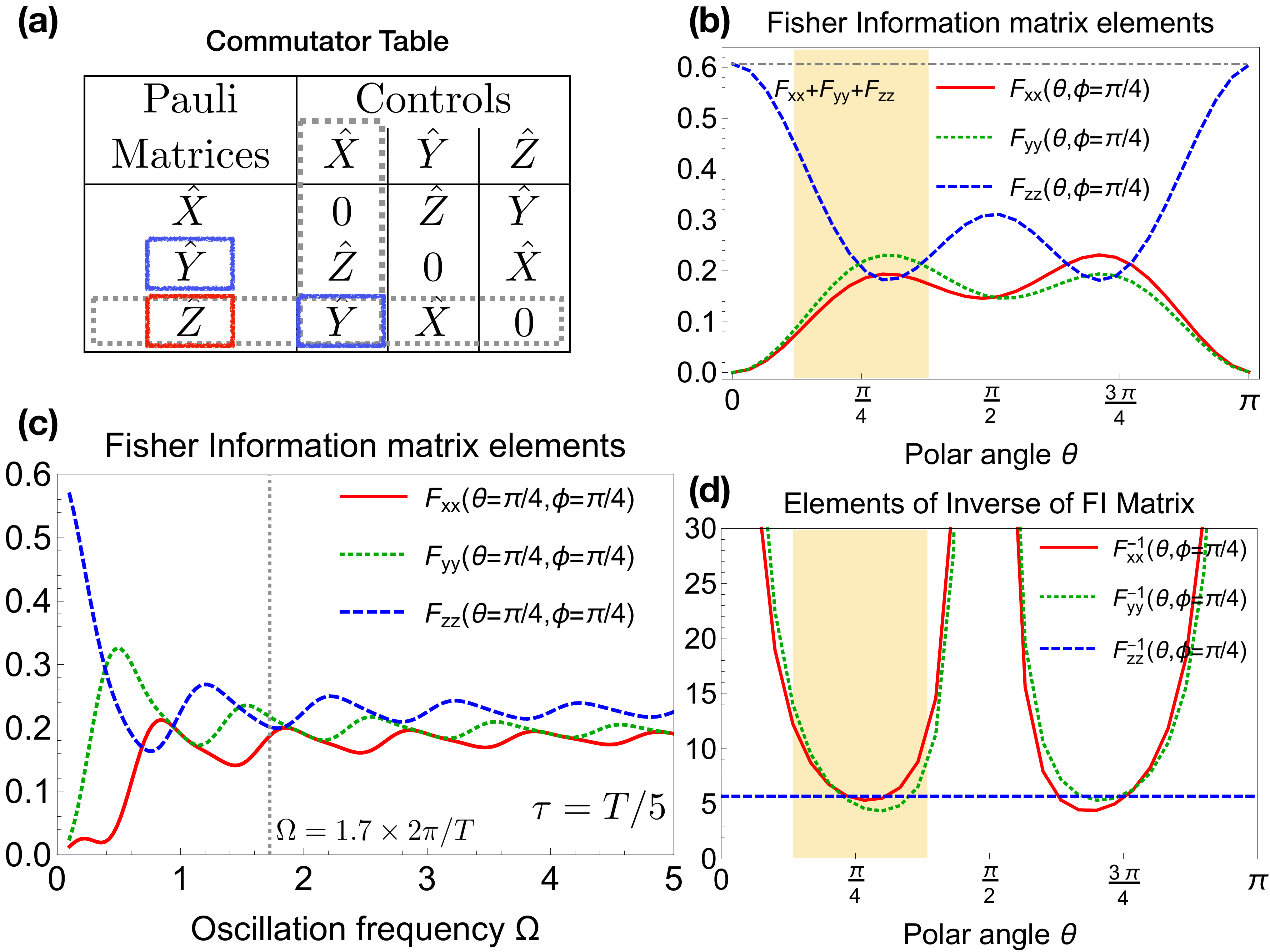}
\caption{Analysis of possible controls for the single-qubit state tomography. (a) The commutator table shows results of all commutators between three observables on the first column and three possible terms in the system's (control) Hamiltonian on the first row. We omit any complex proportional factors of the commutators. The dotted boxes are to highlight the control (${\hat H} \propto \X$) and the measured observable ($\Z$), while the solid boxes are for the observables that can be estimated given the particular control. (b,d) Diagonal elements of the Fisher information matrix and its inverse matrix are plotted with different polar angles $\theta$ of the Rabi oscillation axis. The rotation rate is fixed at $\Omega =  2 \pi (1.7)/T$ where $T$ is the duration of the experiment. The yellow bands show the range of $\theta$ one can use for the qubit tomography. (c) The diagonal elements of the Fisher information matrix with varying $\Omega$, keeping $\theta = \phi = \pi/4$ fixed. The units of $\Omega$ are $2\pi/T$ and we used $\tau = T/5$ for all plots.}
\label{fig-sqbFisher}
\end{figure}

We first use the analysis shown in Eq.~\eqref{eq-Hamevolve} and \eqref{eq-cZtransfer} and construct a commutator table in Fig.~\ref{fig-sqbFisher}(a). The table shows all possible commutators between observables of interest (on the first column) and terms that could exist in the control Hamiltonian (the first row). If one chooses the qubit's oscillation to be around the $x$-axis, i.e., ${\hat H} \propto \X$, then we find that a commutator $[{\hat H}, {\hat E}_{j'} ] \propto \Z$ in \eqref{eq-Hamevolve} has only one possibility which is ${\hat E}_{j'} = \Y$ (shown as the intersection of dashed boxes in Fig.~\ref{fig-sqbFisher}(a)). For the second order term, we find that $[\X, [\X, {\hat E}_{j''}]] \propto \Z$ is satisfied only when ${\hat E}_{j''} = \Z$, which is nothing new. We can conclude that the measured component $c_{\Z}^{\dt}$ in \eqref{eq-cZtransfer} can only be a function of initial state components of $\Z$ and $\Y$, but not $\X$. In other words, only the information about the $\Y$ initial coordinate can be transferred to the measured $\Z$ component via the rotation given by ${\hat H} \propto \X$.

To obtain knowledge of all three qubit components, one can guess that the Rabi control should be a combination of at least two of the three observables. For example, applying a control proportional to $\X + \Y$ should lead to the transfer of information of the $x$ and $y$ qubit coordinates to the observed component of $\Z$. We can write arrow diagrams as,
\begin{align}
\nonumber \X \xrightarrow{\Y} \Z, \\
\nonumber \Y \xrightarrow{\X} \Z,
\end{align}
where the terms in the Hamiltonian responsible for the information transfer are written above the arrows. 

We then compute the Fisher information matrix \eqref{eq-Fishermat2} for the weak measurement limit and show in Fig.~\ref{fig-sqbFisher}(b,c,d) the diagonal elements $F_{xx}$, $F_{yy}$ and $F_{zz}$ for different angles of the Rabi axis and oscillation frequencies. In the panels (b) and (d), we vary the polar angle $\theta$ of the Rabi axis, keeping the azimuthal angle fixed at $\phi = \pi/4$ (in fact this angle does not effect much the estimation quality and can be chosen arbitrary). 
As $\theta$ grows, the information of $z$ decreases from its maximum value, while $F_{xx}$ and $F_{yy}$ increase and reach roughly the same level as $F_{zz}$ at around $\theta  \sim \pi/4 \pm \pi/8$. Fig~\ref{fig-sqbFisher}(c) shows diagonal elements for the inverse of the Fisher information matrix $C \equiv F^{-1}$ which tells us the lower bound for the variances of the estimated parameters. 

We then choose the oscillation axis to be at an angle $\theta = \pi/4$ and see how the Fisher information changes with varying Rabi oscillation rate $\Omega$ in Figure~\ref{fig-sqbFisher}(c). Increasing the frequency results in changing information access from only $\Z$ to all $\X$, $\Y$, and $\Z$, and the trend is practically unchanged after $\Omega$ reaches $\sim 1.0 \times 2\pi/T$. We note that changing $\tau$ (inversely proportional to the measurement strength) only effects the amount of total information gain during the course of measurement time $T$, but does not effect the relative division of information between the three parameters.

\begin{figure}
\includegraphics[width=8.7cm]{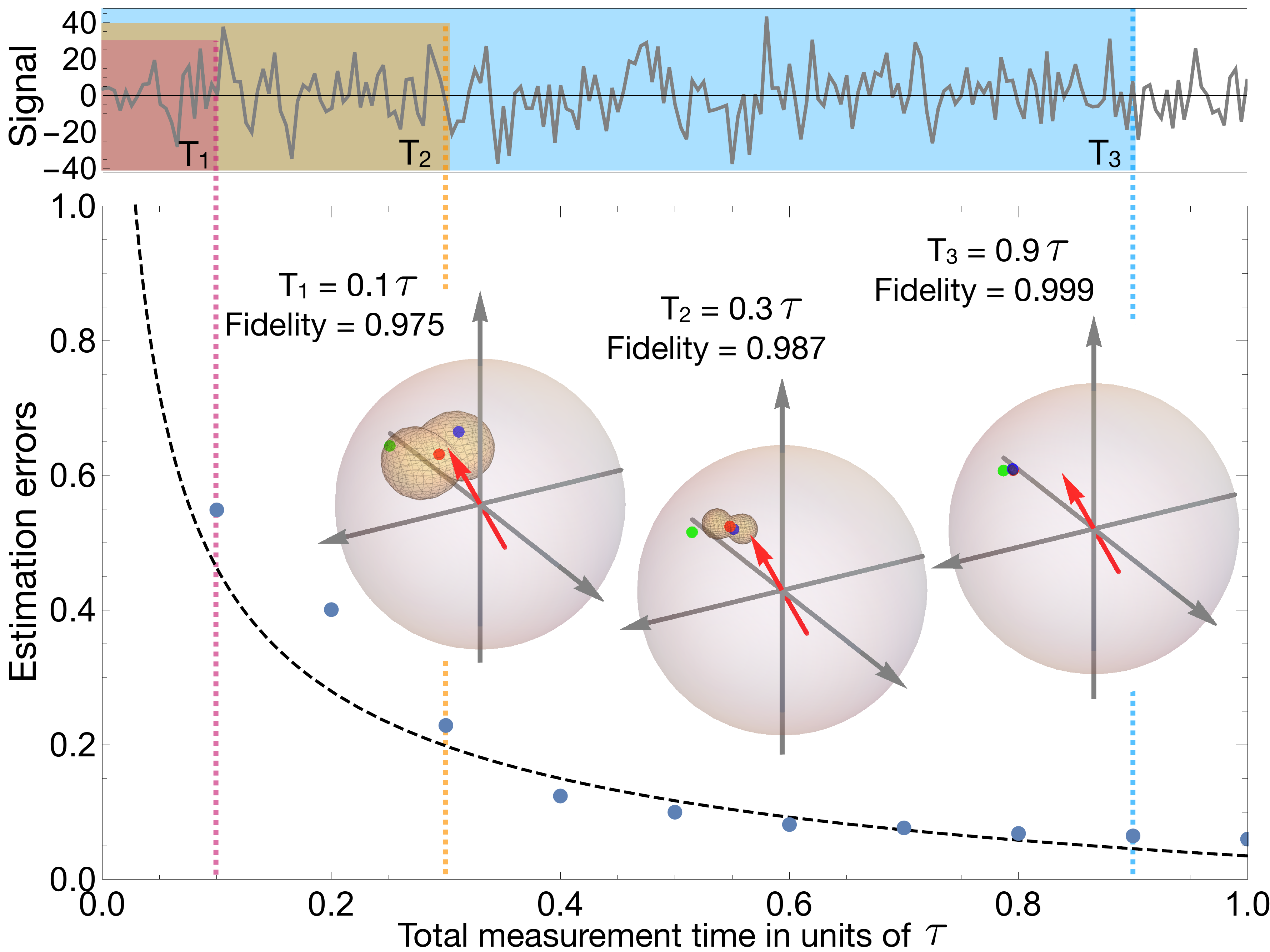}
\caption{Estimation errors of the unknown single-qubit state for different total measurement (data collection) times with $N=5000$. (Top) An example of a measurement record as a function of time. (Bottom, data points) The errors of the estimate calculated from the square root of the trace of the Bayesian covariance matrix Eq.~\eqref{eq-baycovmat}. The fitting curve is $\sim 1/\sqrt{T}$. The three Bloch spheres, at $T_1 = 0.1 \tau$, $T_2 = 0.3 \tau$ and $T_3 = 0.9 \tau$, show the Bayesian estimators (Red), the maximum likelihood estimators (Blue), the unknown true state (Green), and the error ellipsoids from Eq.~\eqref{eq-errorball}. The fidelity between the Bayesian estimators and the true state are also shown for the three measurement times.}
\label{fig-sqbTN}
\end{figure}

To investigate the quality of the state tomography, we numerically simulate quantum trajectories along with their measurement records, and analyze the Bayesian estimators and the maximum likelihood estimators for different measurement and control parameters (see Appendix~\ref{sec-numer} for more detail on the numerical simulation). We consider the mean square error calculated from the trace of a Bayesian covariance matrix, Eq.~\eqref{eq-baycovmat}.
Alternatively, this covariance matrix can be used to construct an \textit{error ellipsoid} for the Bayesian mean estimate. The ellipsoid is described by,
\begin{align}\label{eq-errorball}
r^2= {\vec n}^{\dagger} \cdot C \cdot {\vec n},
\end{align}
where $r$ is the radius of the ellipsoid and is dependent on the direction of a unit vector $\vec n$ pointing out from the Bayesian estimator in the Bloch sphere. 

We show in Figure~\ref{fig-sqbTN} the mean square error of Bayesian estimators which decreases as $1/\sqrt{T}$ where $T$ is the total measurement time. The error ellipsoids, the Bayesian estimators, the maximum likelihood estimators, and the true states are also shown on Bloch spheres at time $T_1 = 0.1 \tau$, $T_2 = 0.3 \tau$ and $T_3 = 0.9 \tau$. As the total time increases, the estimation quality gets better and the size of the error ellipsoid shrinks. The total number of unknown qubits $N$ used in the estimation is chosen to be $5000$ for the data presented in Figure~\ref{fig-sqbTN}, giving the fidelity Eq.~\eqref{eq-fidelity} $F=0.999$ at $T_3$. However, one can improve the fidelity further by increasing $N$. From these results, we learn that the total measurement time $T$ should be at least a few $\tau$ given that $N$ is big enough, so that we can be sure the information about the initial states are collected via the measurement to the desired accuracy. We stress that making the total measurement time $T$ much longer than that will not give further information about the initial state, because the measurement signal becomes uncorrelated with the initial state.

\begin{figure}
\includegraphics[width=8.8cm]{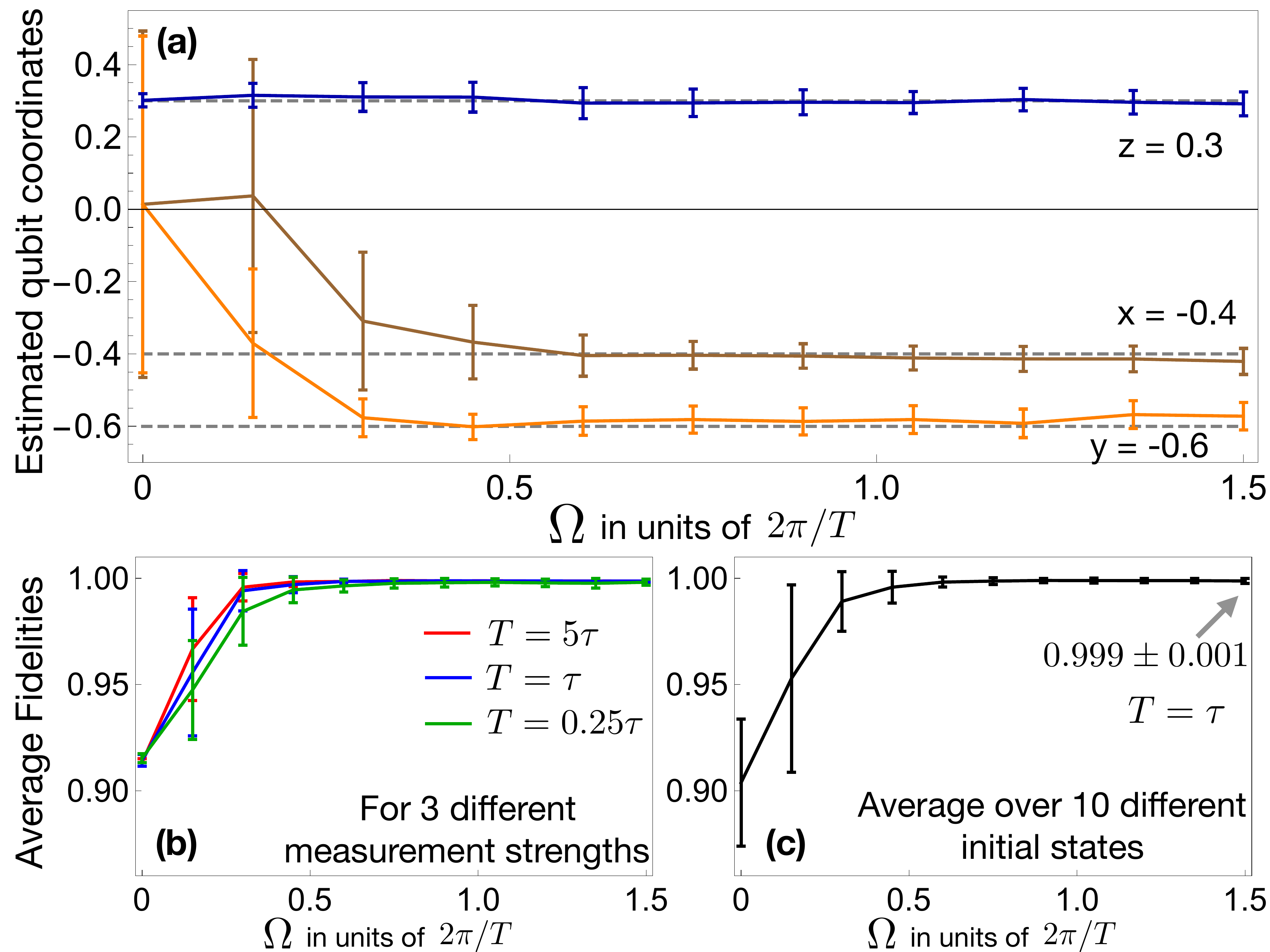}
\caption{Estimated single-qubit coordinates and average fidelities for a fixed Rabi axis ($\theta = \phi = \pi/4$) and different values of the Rabi oscillation rate. (a) The Bayesian estimates of $x,y$, and $z$ coordinates of the unknown true state $\rho_{\rm 0} = (1/2)({\hat I} -0.4 \X -0.6 \Y + 0.3 \Z)$ using $N = 5000$. The error bars of the estimates are calculated from the square root of the trace of the Bayesian covariance matrix Eq.~\eqref{eq-baycovmat}. (b) Average fidelities for 100 simulated Bayesian estimators, for different values of $\tau$ and $\Omega$, where each estimator is computed from $N=5000$ measurement records. The test initial state is the same as in (a). (c) Average fidelities for different initial states chosen randomly (see Appendix~\ref{sec-randomint}). At $\Omega = 2\pi(1.5)/T$, the average fidelity reads $0.999 \pm 0.001$.}
\label{fig-sqbRabi}
\end{figure}

We also perform the numerical simulation for varying Rabi oscillation frequencies, to confirm the trend seen in Fig~\ref{fig-sqbFisher}(c) for the Rabi angle $\theta = \phi = \pi/4$. As expected, see Figure~\ref{fig-sqbRabi}(a), at $\Omega = 0$, the $\Z$ component of the unknown qubit is the only estimated parameter with a very small error bar. As the oscillation frequency reaches $\Omega \sim 0.5 \times 2\pi/T$, the estimation quality for $\X$ and $\Y$ components attain the similar level of precision as the $\Z$ component. One interesting feature is that the estimation error stays unchanged for $\Omega \gtrsim 1.0\times 2\pi/T$, similar to what was found in Fig~\ref{fig-sqbFisher}(c). This can be explained that, because the amount of total information is fixed by $T$, $\tau$ and $N$, the role of the rotation is simply to propagate information of all other qubit coordinates to the measured observable. Once the oscillation rate reaches a threshold $\sim 1.0 \times 2\pi/T$, the estimation quality among all observables stays about the same. 

The fidelity of the state estimation also changes with varying parameters. In Figure~\ref{fig-sqbRabi}(b), the average fidelities for different values of $\tau$ and $\Omega$ are calculated from $100$ numerical data sets, where each set contains one estimate from the Bayesian mean method using $N = 5000$ measurement records. In order to compare the effect of $\tau$ and $\Omega$, the initial unknown state is fixed at the same $\rho_{\rm 0}$. As $\Omega$ increases, the average fidelities increase from the value around 0.91, where only the $z$ coordinate is estimated, to the value close to 1 at a high oscillation rate. The change in $\tau$ results in competing effects between the measurement rate and the oscillation rate, i.e., given fixed $T$, the stronger measurement (red) reaches a high value of state fidelity faster than the weak measurement. However, we note that if the measurement rate is too strong, i.e., much stronger than the oscillation rate $\Omega \ll 1/\tau$, then the measured information will only be about the $\Z$ component, and the fidelity will be bound at $F(\rho_{\rm 0}, (1/2)({\hat I} + 0.3 \Z)) =  0.918$ for this particular $\rho_0$.


In order to obtain the quality of the estimate that is independent of the initial states, we simulate the estimation using 10 randomly chosen initial states. For each one we calculate 100 sets of the Bayesian estimators, and each estimator is calculated from $N = 5000$ measurement records. The average fidelity is shown in the panel (c) of Figure~\ref{fig-sqbRabi} with the error bars computed from the variance of the average. The errors can be understood as coming from within the same initial state (as seen in (b) for example) and from different initial states. The average fidelity reaches $0.999\pm 0.001$ at $\Omega = 1.5 \times 2\pi/T$.

\subsection{Remote-qubit state tomography}\label{sec-mainresults-rqb}
 
 \begin{figure*}
\includegraphics[width=17.5cm]{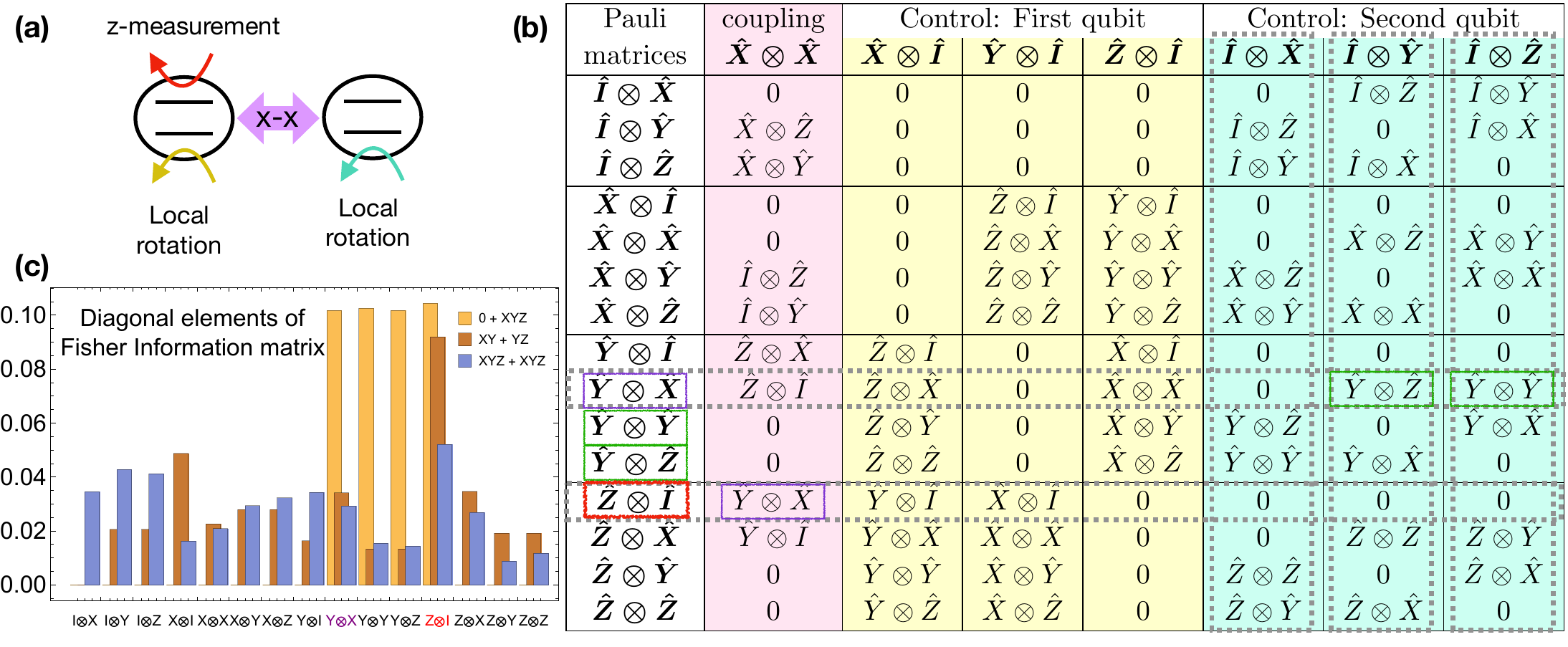}
\caption{Analysis of controls for the remote-qubit and two-qubit state tomography. (a) Schematic diagrams for the two-qubit setup: qubits interact with each other via $\XX$ coupling and Rabi controls are applied locally on each of the qubits. For the remote-qubit tomography, the first qubit is known and acts as an ancilla qubit, while the second qubit is unknown. For the two-qubit tomography, both qubits are unknown. (b) A commutator table shows the commutation relationships between the 15 observables on the first column and 7 terms (terms that can be included in the Hamiltonian of the qubits, Eq.~\eqref{eq-tqbHamil}) in the first row. Any proportional complex constants are omitted. The colored solid boxes show four estimated observables given that the control is in the 0+XYZ setting (see text). (c) Fisher information matrix elements (diagonal elements only) calculated using $g = \Omega = 1.5 \times 2\pi/T$ and $\tau = T/5 $. The colored axis labels in panel (c) are to emphasize the measured observable $\ZI$ and the observable $\YX$ of which its information gets transfer first via the qubit-qubit coupling.}
\label{fig-tqbdiagram}
\end{figure*}

As the second application of our method, we consider quantum state tomography for an unknown single qubit that cannot be measured directly. Instead, the qubit information can only be accessed through an interaction with another qubit. Let us assume that the interaction between the two qubits is described by a $\X\otimes \X$ (capacitive) coupling term, and the possible controls are Rabi drives applied locally on each qubit (see Figure~\ref{fig-tqbdiagram}(a)), with fixed axes and oscillation rates. The two-qubit Hamiltonian for this remote-qubit example (also for the two-qubit example in the next section) is then given by,
\begin{align}\label{eq-tqbHamil}
{\hat H}= \frac{g}{2} \X \otimes \X + \frac{\Omega_{\rm a}}{2} ({\vec n}_{\rm a}\cdot {\hat {\vec \sigma}}) \otimes {\hat I} + \frac{\Omega_{\rm u}}{2} {\hat I} \otimes ({\vec n}_{\rm u} \cdot {\hat {\vec \sigma}}),
\end{align}
where the first term describes the coupling between the two qubits with a coupling strength $g$, and the rest is the local Rabi oscillations with frequencies $\Omega_{\rm a}$ and $\Omega_{\rm u}$, for the \textit{ancilla} qubit and the \textit{unknown} qubit respectively. In this remote-qubit protocol, we assume the first qubit (left one in Fig.~\ref{fig-tqbdiagram}(a)) is a measured qubit with a known prepared state, and the second (right) qubit is an unknown qubit. 

In comparison to the single qubit tomography in the previous subsection, one can think of the coupling as effectively reading out the $\X$ observable of the unknown qubit, as the $\X$ component determines the rotation speed around the $x$-axis of the ancilla qubit, which is then mapped to the measured $\Z$ component. Therefore, a straightforward option for qubit controls is to set the Rabi axis for the unknown qubit at $\theta_{\rm u} = \phi_{\rm u} = \pi/4$, as we found in the single qubit example, and then initialize the ancilla qubit state at $y_{\rm a}^0 = {\rm Tr}(\rho_0 \, \Y \otimes \I)= 1$ with no Rabi drive $\Omega_{\rm a} = 0$. 
 
We can see that this straightforward protocol works by looking at the commutator table and the Fisher information matrix elements shown in Fig~\ref{fig-tqbdiagram}(b) and (c), respectively. If we choose the Rabi axis for the unknown qubit to be in the direction described by $\theta_{\rm u} = \phi_{\rm u} = \pi/4$ and no oscillation on the ancilla qubit, then the Hamiltonian is ${\hat H} = (g/2) \XX + (\Omega_{\rm u}/4)\left( \IX + \IY +\sqrt{2}\, \IZ \right)$. Using the analysis described in Eq.~\eqref{eq-Hamevolve} and \eqref{eq-cZtransfer}, also explained in Figure~\ref{fig-tqbdiagram} and its caption, we find that the state components that can be transferred to the measured observable $\Z \otimes {\hat I}$ via this unitary dynamics are: $\Y \otimes \X$, $\Y \otimes \Y$, and $\Y \otimes \Z$. We can write diagrams to explain the information transfer as,
\begin{align}
\YX \xrightarrow{\XX} \ZI, \\
\YY \xrightarrow{\IZ} \YX \xrightarrow{\XX} \ZI, \\
\YZ \xrightarrow{\IY} \YX \xrightarrow{\XX} \ZI,
\end{align}
which can be interpreted as: the $\YX$ component of the qubits is transferred to the measured $\ZI$ component via the $\XX$ coupling, and the component of $\YY$ (or $\YZ$) is transferred to $\ZI$ via the control term $\IZ$ (or $\IY$) as well as the coupling term. These three observables $\YX$, $\YY$ and $\YZ$ are sufficient for the estimation of three coordinates of the unknown qubit, given that the $y$ coordinate of the ancilla state is initialized in a non-zero value $y_{\rm a}^0 \ne 0$. The analysis of the Fisher information matrix in the weak measurement limit also gives non-zero values for the three observables (in addition to the measured $\ZI$). This is shown as the yellow bars for ``0+XYZ'' setting in Fig~\ref{fig-tqbdiagram}(c). For convenience, we use the code ``A+B'' to represent a set of Rabi controls, where A is a combination of X, Y, and Z describing a Rabi axis of the first qubit with non-zero components in $\X$, $\Y$, and $\Z$ (the same goes for B for the second qubit's rotation axis). If A = 0, it refers to no Rabi oscillation on the first qubit.

We numerically simulate the tomography process to test the quality of the estimation using the proposed 0+XYZ setting. Similar to the previous single qubit example, we compute a Bayesian estimated state from the probabilities of the trial states given the measurement signals (see Appendix~\ref{sec-numer} for the detail on the numerical simulation). To analyze the quality of the estimation, we simulate $100$ estimators (each estimator is computed from $N=5000$ measurement records) for each set of the control parameters of interest and compute root mean square errors (RMSEs) from the $100$ estimators compared with a given true state. 

\begin{figure}
\includegraphics[width=8.6cm]{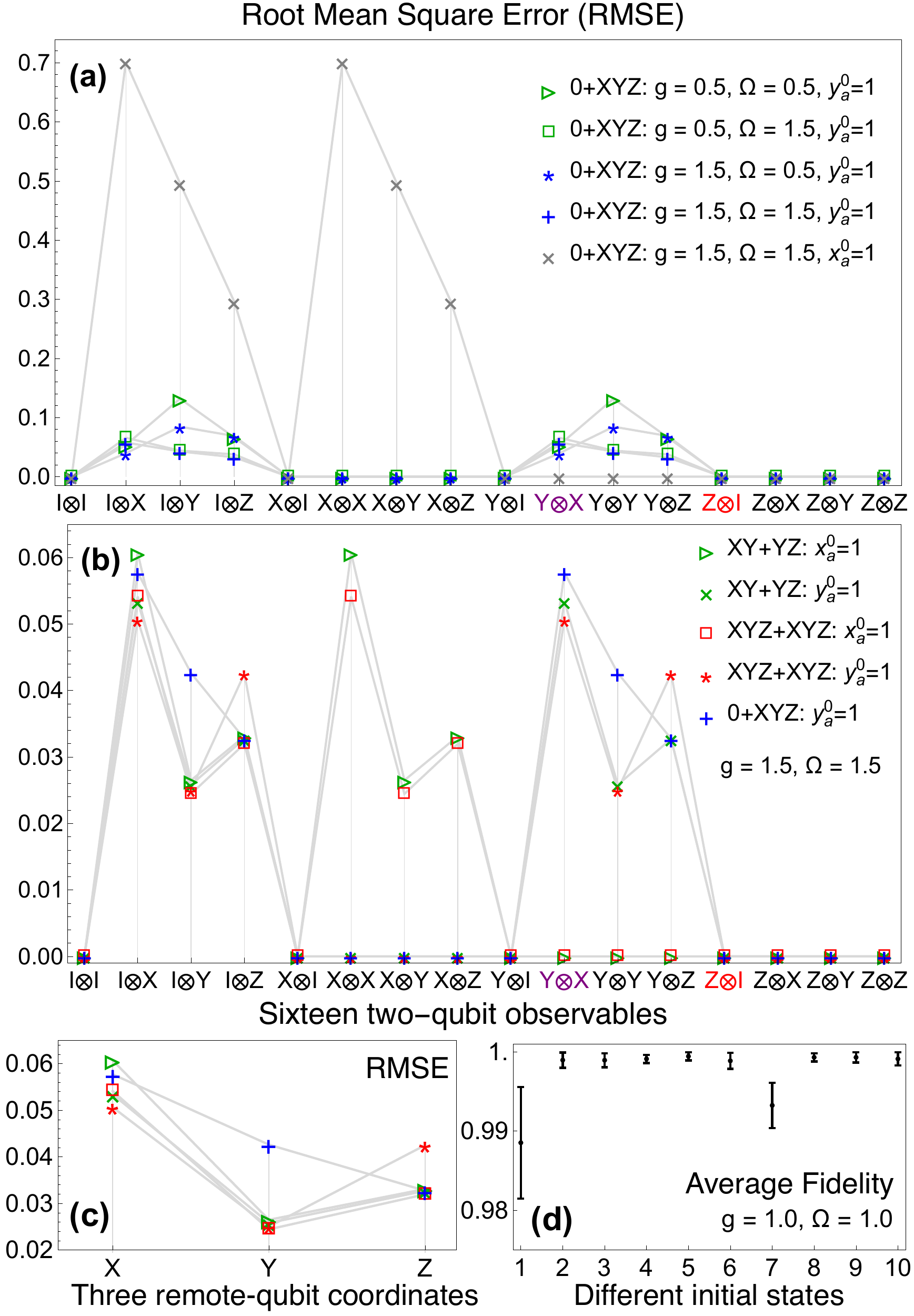}
\caption{Root mean square errors for the remote qubit tomography. (a) shows the errors of the estimated 16 remote-and-ancilla-qubit coordinates for the control settings where the Rabi oscillation is only applied on the remote qubit, with different values of $g$, $\Omega_{\rm u} = \Omega$, and the initial states of the ancilla qubit (denoted by $x_a^0,y_a^0$). (b,c) shows the estimation errors for successful control settings using the coupling strength $g = 1.5$ and the qubit oscillation frequency $\Omega_{\rm a} = \Omega_{\rm u} = \Omega = 1.5$. (b) shows the errors of all 16 elements, while (c) shows only the errors of estimated remote-qubit coordinates. Note that the vertical scale of (a) is an order of magnitude lower than (b). The abbreviations 0+XYZ, XY+YZ, etc., denote the controls and are defined in the text. Data points in (a,b,c) are numerically simulated with the same unknown remote qubit state $\rho_{\rm u} = (1/2)(\I + 0.7 \X -0.5 \Y +0.3\Z)$ using $N=5000$ and the same measurement strength $\tau = T/5 $. We test the remote-qubit state estimation with different initial states using the XY+YZ setting with $g = \Omega = 1.0$ and show average fidelities with their error bars in (d). The units of $g$ and $\Omega$ are $2 \pi/T$.}
\label{fig-rqbresults}
\end{figure}

The results are shown in Fig~\ref{fig-rqbresults}(a), presenting the RMSEs for the estimation of the 16 two-qubit density matrix components. Large values of RMSE mean that the elements are poorly estimated, or simply that little information about them was obtained from the measurement records. The vanishing errors, on the other hand, represent absolute accuracy, which in this case happens to the elements that are known to have zero values. For example, choosing $y_{\rm a}^0= 1$ (meaning that $x_{\rm a}^0= z_{\rm a}^0 = 0$), we know with certainty that the elements other than $\IX, \IY, \IZ, \YX, \YY$, and $\YZ$ of the initial state have zero values. From Figure~\ref{fig-rqbresults}(a), for the 0+XYZ setting, the RMSEs tell us that the quality of the estimation is significantly better when $y_{\rm a}^0 = 1$ than when $x_{\rm a}^0=1$, and can be even better if $g$ and $\Omega$ are in the range of $1.0 - 1.5$ in units of $2\pi/T$. In Fig.~\ref{fig-rqbresults}(b,c), we show the errors for the 0+XYZ setting using $y_{\rm a}^0 = 1$ and the other two settings used in the full two-qubit tomography (discussed in the next section). This is to show that the controls that give access to all 15 two-qubit components can also be used to estimate the 3 components of the remote unknown qubit with comparable precision. We show average fidelities for various unknown states in Fig.~\ref{fig-rqbresults}(d), where the ten randomly chosen unknown states are listed in the Appendix~\ref{sec-randomint}. The average fidelities stay in the range of $0.99\pm 0.01$ for $N= 5000$.




\subsection{Two-qubit state tomography}\label{sec-mainresults-tqb}

As the third application, we consider a full two-qubit tomography where all of the 15 observables of the coupled two qubits can be accessed by only continuously measuring the observable $\ZI$ and applying the fixed qubit controls. From the commutator table in Fig~\ref{fig-tqbdiagram}(b), one can convince oneself that a ``XY+YZ'' control, where ${\vec n}_{\rm a} = (1/\sqrt{2})(\X + \Y)$ and ${\vec n}_{\rm u} = (1/\sqrt{2}) (\Y + \Z)$, and a more overall ``XYZ+XYZ'' control, where ${\vec n}_{\rm a} =(1/2)( \X + \Y +\sqrt{2} \Z)$ and ${\vec n}_{\rm u} = (1/\sqrt{2})(\X + \Y +\sqrt{2} \Z)$, lead to accessing information of all 15 observables. The Fisher information matrix analysis for the two settings are shown in Figure~\ref{fig-tqbdiagram}(c). In this subsection, we use these sets of controls to examine their performance in estimating full unknown two-qubit states, including pure states, mixed states, and non-separable states. 


\begin{figure}
\includegraphics[width=8.6cm]{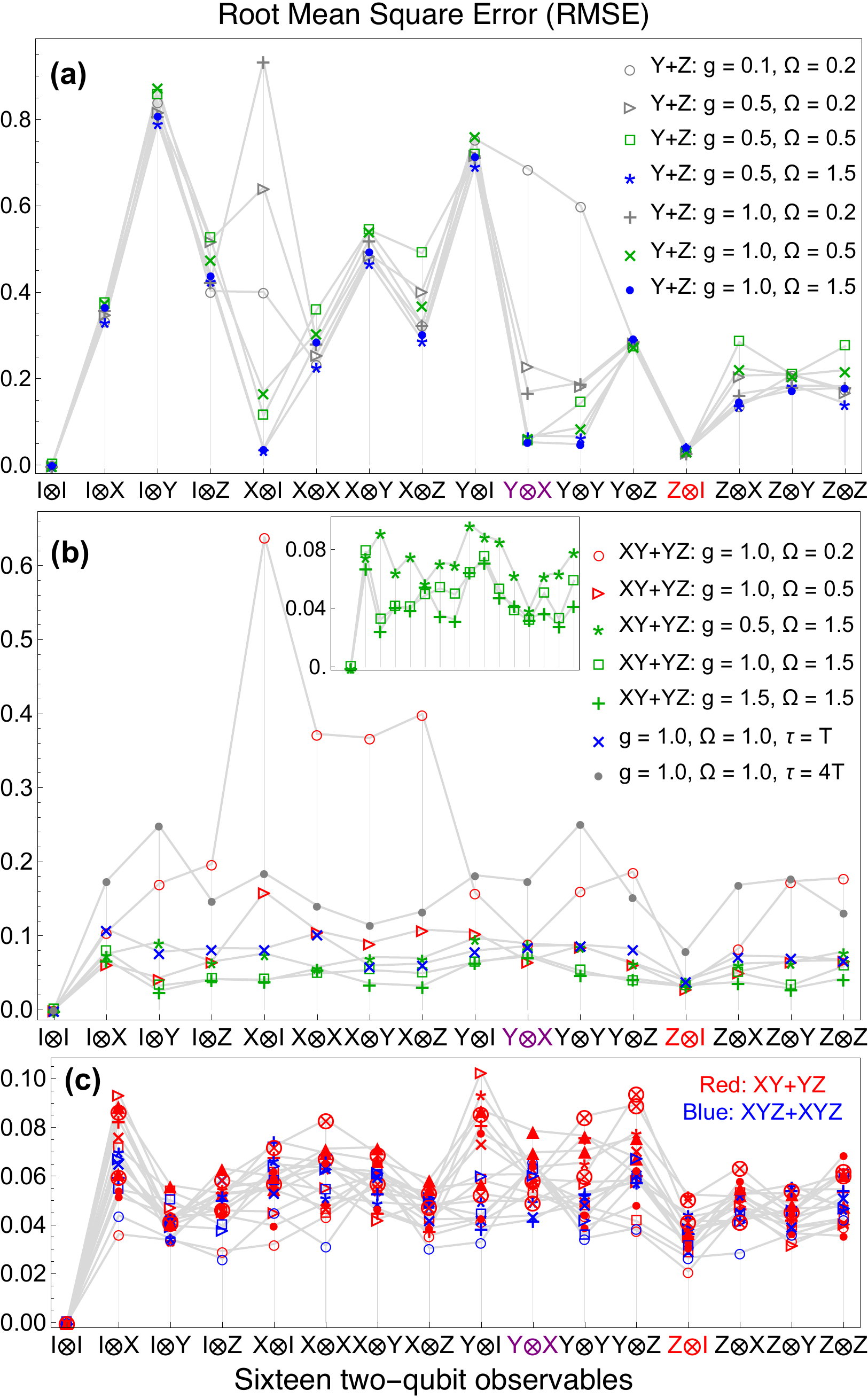}
\caption{Root mean square errors for the full two-qubit state tomography. (a) shows the estimation errors for the Y+Z control setting with different values of $g$ and $\Omega$. The initial state is fixed at a product state between $\rho_1 = (1/2)(\I - 0.4 \X  -0.75 \Y + 0.5 \Z)$ and $\rho_2 = (1/2)(\I + 0.6 \X -0.5\Y + 0.6 \Z)$, where $\tau = T/5$, and $N=4000$. The RMSEs are calculated from $50$ sets of Maximum Likelihood estimators for each set of parameters. (b) shows the estimation errors for the XY+YZ control, with the same initial unknown state and number of sets as in (a). We use $\tau = T/5$ except for last two sets of data, where we use $\tau = T$ and $\tau = 4 T$. The inset shows small improvement from increasing $g$ and keeping $\Omega = 1.5\times 2\pi/T$. (c) shows the errors for the two successful settings, XY+YZ and XYZ+XYZ, for 9 different initial states for each setting, using $g = \Omega = 1.5 \times 2\pi/T$ and $\tau = T/5$. The average fidelity of all 18 sets is $0.984 \pm 0.003$.}
\label{fig-tqbresults}
\end{figure}

In contrast to the three dimensional real-number space of the Bloch sphere representing single-qubit states in the previous two examples, the state space of the two-qubit system, is a space of 15 real numbers with finite ranges. This large state space makes the Bayesian estimation a non-preferable method, as one needs to calculate the posterior probability density from Eq~\eqref{eq-likelihoodR} for each of candidate states. To get a similar accuracy as in the single-qubit case, we need much more than $10^8$ trial two-qubit states distributed uniformly with a specific measure (or with some prior distribution), only to compute one Bayesian estimator (see more detail in Appendix~\ref{sec-numer}). Therefore, to minimize the computation overhead, we instead maximize the likelihood function Eq.~\eqref{eq-likelihoodR} over the valid state space using a search algorithm. The estimator in this two-qubit tomography protocol is then the maximum likelihood estimator defined in Eq.~\eqref{eq-MLE}.

To guarantee valid physical quantum states of the MLE, we need to specify a real-number space to be used with a numerical search algorithm. For an expansion of components of a two-qubit state given by,
\begin{align}
\nonumber \rho= & \frac{1}{4}\sum_{i,j}\ccij[i][j]{\hat \sigma}_i \otimes {\hat \sigma}_j,
\end{align}
where the indices are $i,j = 0,1,2,3$, and the Pauli matrices ${\hat \sigma}_j$ for $j = 0$, 1, 2, and 3, represent $\I$, $\X$, $\Y$, and $\Z$, respectively. We then have $c_{00} = 1$ and the rest are real numbers.  Following the calculation shown in \cite{Omar2016}, if we write the coefficients $\ccij$ in a matrix form,
\begin{equation}
D=\begin{pmatrix}1 & \ccij[0][1] & \ccij[0][2] & \ccij[0][3]\\
\ccij[1][0] & \ccij[1][1] & \ccij[1][2] & \ccij[1][3]\\
\ccij[2][0] & \ccij[2][1] & \ccij[2][2] & \ccij[2][3]\\
\ccij[3][0] & \ccij[3][1] & \ccij[3][2] & \ccij[3][3]
\end{pmatrix}\equiv\begin{pmatrix}1 & \vv^{\dagger}\\
\vu & B
\end{pmatrix},
\end{equation}
where we have defined the vector $\vv$, $\vu$, and the matrix $B$ on the right hand side, the positivity condition for the two-qubit state is given by three constraints,
\begin{align}\label{eq-constraints}
0\leq & \,\, 4-\|D\|^{2},\\
0 \leq & \,\, 2(\vu^{\dagger}B\vv-\det B)-(\|D\|^{2}-2),\\
\nonumber 0 \leq & \, \, 8(\vu^{\dagger}B\vv-\det B) +(\|D\|^{2}-2)^{2}+8\vu^{\dagger} {\tilde B} \vv\\
& -4(\|u\|^{2}\|v\|^{2}+\|\vu^{\dagger}B\|^{2}+\|B\vv\|^{2}+\| {\tilde B} \|^{2}),
\end{align}
where ${\tilde B}$ is the cofactor matrix of $B$, and $\| A \| = {\rm Tr}[A^{\dagger} A]$ using the Hilbert-Schmidt inner product. We note that, for the numerical search, we use the Differential Evolution algorithm to find a state that maximizes Eq.~\eqref{eq-likelihoodR}, with $50$ initial random searches to reach a reasonable convergence (more detail of the numerical method in Appendix~\ref{sec-numer}).

For comparison purpose, we first show numerical results for the two-qubit state tomography for a poorly chosen set of controls using a control setting ``Y+Z,'' where the Rabi oscillations of the first and second qubits are around the $y$-axis and $z$-axis, respectively. The RMSEs for this setting is shown in Fig~\ref{fig-tqbresults}(a) for different values of control parameters $g$ and $\Omega$. Only three components for $\XI$, $\YX$, and $\YY$ that show significant improvement in the errors, which are reduced to less than $0.1$ while increasing the parameters $g$ and $\Omega$ to be around $\sim 1.0 - 1.5$, in units of $2\pi/T$. This agrees with a prediction from the two-qubit commutator table in Fig.~\ref{fig-tqbdiagram}(b), where one can deduce that the unitary dynamics allow access to the information of components for only four observables: $\XI, \YX, \YY$, and  $\ZI$. 



To estimate all the 15 elements of the unknown two-qubit state concurrently, we use the control settings: XY+YZ or XYZ+XYZ, mentioned in the beginning of this section. Since the XY+YZ control is simpler to numerically simulate than the latter, we choose this control as our preferred setting; however, the results are similar for the XYZ+XYZ control. We show in Figure~\ref{fig-tqbresults}(b) the estimation errors for different control parameter values. As the coupling strength and the oscillation frequency grow, the estimation errors for all 15 elements are significantly reduced to values less than $0.1$, at around $g \sim 1.0 -1.5$ and $\Omega \sim 1.0-1.5$ in units of $2\pi/T$. In the plot, we also show the effect of the measurement strength by changing $\tau$ keeping the controls fixed, as seen in the last two data sets of the figure legend. As expected, weakening the measurement strength simply lifts up the baseline of the errors of all 15 elements, which can be interpreted that the relative information distributed among them is approximately the same, and only the total information of all parameters changes. 

From these numerical results in Figure~\ref{fig-tqbresults}(b), we find that the quality of the estimation is well saturated at around $g \sim \Omega \sim 1.0-1.5$ in units of $2\pi/T$, similar ranges to what we have found in the previous single-qubit and remote-qubit cases. We then perform the numerical tests for different randomly-chosen unknown states, containing product states and non-separable state mixtures (e.g., Werner states) with varied purities, using $g = \Omega = 1.5 \times 2\pi/T$ for both XY+YZ and XYZ+XYZ settings. As shown in Figure~\ref{fig-tqbresults}(c), the RMSEs for nine different initial states per one setting are roughly the same, varying in the small range of $0.02$ to $0.1$. We also calculate the average fidelity for all these eighteen cases giving a value of $0.984$ with an error bar of $0.003$. We note that these results are from maximum likelihood estimators using only $N = 4000$, and thus the fidelity should be better if one have larger numbers of $N$ used in the estimation.

We note that both XY+YZ and XYZ+XYZ settings can also be used for the remote qubit tomography, since the information about the unknown qubit coordinates is embedded in the components of all twelve observables: $\IX, \IY, \IZ, ..., \ZY, \ZZ$, given that the initial state of the ancilla qubit is known. We  show the results of numerical simulation in Figure~\ref{fig-rqbresults}(b). Both settings perform equally well independent of the initial ancilla states ($x_{\rm a}^0 = 1$, $y_{\rm a}^0 = 1$, or a mixture of non-zero $x_{\rm a}^0$ and $y_{\rm a}^0$). 


\section{Discussion and conclusion}\label{sec-conclusion}

We have analyzed our proposed method for quantum state tomography using continuous weak measurement where a system of qubits with an unknown state is measured with resource limitation. We consider in particular a system of interacting qubits that can be measured only through one of its observables, and the qubit controls are simple Rabi oscillation applied locally on each qubit. We find measurement and control settings, consisting of values of the measurement strength, the oscillation axes, the oscillation rates, and the coupling rates, that allows access to the information of all unknown components of qubit observables in one measurement run. 

The proposed tomographic scheme is analyzed using three proof-of-principle applications: (a) single qubit, (b) remote qubit, and (c) full two qubit tomography, with varying control settings and estimation methods (Bayesian mean and maximum likelihood estimation). The control settings are chosen based on the analysis of the commutation relations as well as the Fisher information matrix of the estimated components. In the single-qubit tomography, a continuous probe of the $\Z$ observable and a Rabi oscillation with an axis of rotation off-parallel from the measured $z$-axis are sufficient to extract information of the three qubit coordinates. For the remote-qubit and the full two-qubit tomography, a continuous measurement of the $\ZI$ observable is sufficient to extract the qubits' information if it is combined with specific local qubit controls. The axes of local qubit controls we found most efficient are: (a) the rotation axis of the measured qubit is aligned diagonally in $x-y$ plane and the rotation axis of the unmeasured qubit is aligned diagonally in $y-z$ plane (denoted as XY+YZ setting), and (b) the rotation axes of both qubits are off-parallel from all $x,y,z$ axes (denoted as XYZ+XYZ setting).

Once the axes of the oscillations are fixed, from our analysis, we found that the rotation rates need to be at least $1.0\times 2\pi/T$, in order to assure uniformly extracted information of all qubits' components. In other words, the oscillation rates should be fast enough that a net rotated angle (ignoring the measurement backaction) covers at least a full $2\pi$ rotation during the total measurement time $T$. The coupling rate of the two qubits is found to be optimal at the same value as the local Rabi oscillation rate. Moreover, the measurement strength should be strong enough that all information is collected from the measurement, but not too strong that it results in solely extracting the information of the measured observable and less information about other observables. That is, the total measurement time should to about one or two times the characteristic measurement time $\tau$.

As a summary, we have demonstrated that quantum state tomography can be achieved using only limited measurement and fixed control resources, if the control settings are chosen wisely. The reason why this scheme works can be explained as follows. The unitary dynamics and the measurement backaction play the role of a random sampler, effectively mapping the measured observable (in this case is the $\Z$ observable of a qubit) to 
a series of random observables that can sample the unknown state in a variety of different bases continuously in time. 
We stress the relative simplicity of this method, which should be practical for implementation in the laboratory.
As an outlook for future work, it is natural to investigate the optimality of the methods, and their state estimation performance in comparison to other conventional methods, such as the linear inversion, taking into account the practical finite-strength quantum measurement. Moreover, the issue of optimality of these tomographic schemes also poses many interesting geometrical questions whether they can achieve a uniform measurement sampling of a quantum state.

\begin{acknowledgments}
This work was supported by US Army Research Office Grant No. W911NF-15-1-0496, and by the National Science Foundation grant DMR-1506081. AC acknowledges support by the Australian Research Council Centre of Excellence CE170100012. TC acknowledges the Sri Trang Thong scholarship for the undergraduate program at Mahidol University and for the summer research at University of Rochester. 
\end{acknowledgments}

\bibliographystyle{apsrev4-1}

\begin{thebibliography}{41}%
\makeatletter
\providecommand \@ifxundefined [1]{%
 \@ifx{#1\undefined}
}%
\providecommand \@ifnum [1]{%
 \ifnum #1\expandafter \@firstoftwo
 \else \expandafter \@secondoftwo
 \fi
}%
\providecommand \@ifx [1]{%
 \ifx #1\expandafter \@firstoftwo
 \else \expandafter \@secondoftwo
 \fi
}%
\providecommand \natexlab [1]{#1}%
\providecommand \enquote  [1]{``#1''}%
\providecommand \bibnamefont  [1]{#1}%
\providecommand \bibfnamefont [1]{#1}%
\providecommand \citenamefont [1]{#1}%
\providecommand \href@noop [0]{\@secondoftwo}%
\providecommand \href [0]{\begingroup \@sanitize@url \@href}%
\providecommand \@href[1]{\@@startlink{#1}\@@href}%
\providecommand \@@href[1]{\endgroup#1\@@endlink}%
\providecommand \@sanitize@url [0]{\catcode `\\12\catcode `\$12\catcode
  `\&12\catcode `\#12\catcode `\^12\catcode `\_12\catcode `\%12\relax}%
\providecommand \@@startlink[1]{}%
\providecommand \@@endlink[0]{}%
\providecommand \url  [0]{\begingroup\@sanitize@url \@url }%
\providecommand \@url [1]{\endgroup\@href {#1}{\urlprefix }}%
\providecommand \urlprefix  [0]{URL }%
\providecommand \Eprint [0]{\href }%
\providecommand \doibase [0]{http://dx.doi.org/}%
\providecommand \selectlanguage [0]{\@gobble}%
\providecommand \bibinfo  [0]{\@secondoftwo}%
\providecommand \bibfield  [0]{\@secondoftwo}%
\providecommand \translation [1]{[#1]}%
\providecommand \BibitemOpen [0]{}%
\providecommand \bibitemStop [0]{}%
\providecommand \bibitemNoStop [0]{.\EOS\space}%
\providecommand \EOS [0]{\spacefactor3000\relax}%
\providecommand \BibitemShut  [1]{\csname bibitem#1\endcsname}%
\let\auto@bib@innerbib\@empty
\bibitem [{\citenamefont {Newton}\ and\ \citenamefont
  {Young}(1968)}]{Newton1968}%
  \BibitemOpen
  \bibfield  {author} {\bibinfo {author} {\bibfnamefont {R.~G.}\ \bibnamefont
  {Newton}}\ and\ \bibinfo {author} {\bibfnamefont {B.-l.}\ \bibnamefont
  {Young}},\ }\href@noop {} {\bibfield  {journal} {\bibinfo  {journal} {Ann.
  Phys.}\ }\textbf {\bibinfo {volume} {49}},\ \bibinfo {pages} {393} (\bibinfo
  {year} {1968})}\BibitemShut {NoStop}%
\bibitem [{\citenamefont {Nielsen}\ and\ \citenamefont
  {Chuang}(2000)}]{Nielsen2000}%
  \BibitemOpen
  \bibfield  {author} {\bibinfo {author} {\bibfnamefont {M.~A.}\ \bibnamefont
  {Nielsen}}\ and\ \bibinfo {author} {\bibfnamefont {I.~L.}\ \bibnamefont
  {Chuang}},\ }\href@noop {} {\emph {\bibinfo {title} {Quantum {{Computation}}
  and {{Quantum Information}}}}}\ (\bibinfo  {publisher} {{Cambridge University
  Press}},\ \bibinfo {address} {Cambridge; New York},\ \bibinfo {year}
  {2000})\BibitemShut {NoStop}%
\bibitem [{\citenamefont {James}\ \emph {et~al.}(2001)\citenamefont {James},
  \citenamefont {Kwiat}, \citenamefont {Munro},\ and\ \citenamefont
  {White}}]{Jamestomo2001}%
  \BibitemOpen
  \bibfield  {author} {\bibinfo {author} {\bibfnamefont {D.~F.~V.}\
  \bibnamefont {James}}, \bibinfo {author} {\bibfnamefont {P.~G.}\ \bibnamefont
  {Kwiat}}, \bibinfo {author} {\bibfnamefont {W.~J.}\ \bibnamefont {Munro}}, \
  and\ \bibinfo {author} {\bibfnamefont {A.~G.}\ \bibnamefont {White}},\ }\href
  {\doibase 10.1103/PhysRevA.64.052312} {\bibfield  {journal} {\bibinfo
  {journal} {Phys. Rev. A}\ }\textbf {\bibinfo {volume} {64}},\ \bibinfo
  {pages} {052312} (\bibinfo {year} {2001})}\BibitemShut {NoStop}%
\bibitem [{\citenamefont {de~Burgh}\ \emph {et~al.}(2008)\citenamefont
  {de~Burgh}, \citenamefont {Langford}, \citenamefont {Doherty},\ and\
  \citenamefont {Gilchrist}}]{Doherty2008}%
  \BibitemOpen
  \bibfield  {author} {\bibinfo {author} {\bibfnamefont {M.~D.}\ \bibnamefont
  {de~Burgh}}, \bibinfo {author} {\bibfnamefont {N.~K.}\ \bibnamefont
  {Langford}}, \bibinfo {author} {\bibfnamefont {A.~C.}\ \bibnamefont
  {Doherty}}, \ and\ \bibinfo {author} {\bibfnamefont {A.}~\bibnamefont
  {Gilchrist}},\ }\href {\doibase 10.1103/PhysRevA.78.052122} {\bibfield
  {journal} {\bibinfo  {journal} {Phys. Rev. A}\ }\textbf {\bibinfo {volume}
  {78}},\ \bibinfo {pages} {052122} (\bibinfo {year} {2008})}\BibitemShut
  {NoStop}%
\bibitem [{\citenamefont {Gross}\ \emph {et~al.}(2010)\citenamefont {Gross},
  \citenamefont {Liu}, \citenamefont {Flammia}, \citenamefont {Becker},\ and\
  \citenamefont {Eisert}}]{Grosssensing2010}%
  \BibitemOpen
  \bibfield  {author} {\bibinfo {author} {\bibfnamefont {D.}~\bibnamefont
  {Gross}}, \bibinfo {author} {\bibfnamefont {Y.-K.}\ \bibnamefont {Liu}},
  \bibinfo {author} {\bibfnamefont {S.~T.}\ \bibnamefont {Flammia}}, \bibinfo
  {author} {\bibfnamefont {S.}~\bibnamefont {Becker}}, \ and\ \bibinfo {author}
  {\bibfnamefont {J.}~\bibnamefont {Eisert}},\ }\href {\doibase
  10.1103/PhysRevLett.105.150401} {\bibfield  {journal} {\bibinfo  {journal}
  {Phys. Rev. Lett.}\ }\textbf {\bibinfo {volume} {105}},\ \bibinfo {pages}
  {150401} (\bibinfo {year} {2010})}\BibitemShut {NoStop}%
\bibitem [{\citenamefont {Renes}\ \emph {et~al.}(2004)\citenamefont {Renes},
  \citenamefont {Blume-Kohout}, \citenamefont {Scott},\ and\ \citenamefont
  {Caves}}]{RenessicPOVM2016}%
  \BibitemOpen
  \bibfield  {author} {\bibinfo {author} {\bibfnamefont {J.~M.}\ \bibnamefont
  {Renes}}, \bibinfo {author} {\bibfnamefont {R.}~\bibnamefont {Blume-Kohout}},
  \bibinfo {author} {\bibfnamefont {A.~J.}\ \bibnamefont {Scott}}, \ and\
  \bibinfo {author} {\bibfnamefont {C.~M.}\ \bibnamefont {Caves}},\ }\href
  {\doibase 10.1063/1.1737053} {\bibfield  {journal} {\bibinfo  {journal}
  {Journal of Mathematical Physics}\ }\textbf {\bibinfo {volume} {45}},\
  \bibinfo {pages} {2171} (\bibinfo {year} {2004})},\ \Eprint
  {http://arxiv.org/abs/https://doi.org/10.1063/1.1737053}
  {https://doi.org/10.1063/1.1737053} \BibitemShut {NoStop}%
\bibitem [{\citenamefont {Bassa}\ \emph {et~al.}(2015)\citenamefont {Bassa},
  \citenamefont {Goyal}, \citenamefont {Choudhary}, \citenamefont {Uys},
  \citenamefont {Di\'osi},\ and\ \citenamefont {Konrad}}]{Bassa2015}%
  \BibitemOpen
  \bibfield  {author} {\bibinfo {author} {\bibfnamefont {H.}~\bibnamefont
  {Bassa}}, \bibinfo {author} {\bibfnamefont {S.~K.}\ \bibnamefont {Goyal}},
  \bibinfo {author} {\bibfnamefont {S.~K.}\ \bibnamefont {Choudhary}}, \bibinfo
  {author} {\bibfnamefont {H.}~\bibnamefont {Uys}}, \bibinfo {author}
  {\bibfnamefont {L.}~\bibnamefont {Di\'osi}}, \ and\ \bibinfo {author}
  {\bibfnamefont {T.}~\bibnamefont {Konrad}},\ }\href {\doibase
  10.1103/PhysRevA.92.032102} {\bibfield  {journal} {\bibinfo  {journal} {Phys.
  Rev. A}\ }\textbf {\bibinfo {volume} {92}},\ \bibinfo {pages} {032102}
  (\bibinfo {year} {2015})}\BibitemShut {NoStop}%
\bibitem [{\citenamefont {Silberfarb}\ \emph {et~al.}(2005)\citenamefont
  {Silberfarb}, \citenamefont {Jessen},\ and\ \citenamefont
  {Deutsch}}]{SilberfarbPRL2005}%
  \BibitemOpen
  \bibfield  {author} {\bibinfo {author} {\bibfnamefont {A.}~\bibnamefont
  {Silberfarb}}, \bibinfo {author} {\bibfnamefont {P.~S.}\ \bibnamefont
  {Jessen}}, \ and\ \bibinfo {author} {\bibfnamefont {I.~H.}\ \bibnamefont
  {Deutsch}},\ }\href {\doibase 10.1103/PhysRevLett.95.030402} {\bibfield
  {journal} {\bibinfo  {journal} {Phys. Rev. Lett.}\ }\textbf {\bibinfo
  {volume} {95}},\ \bibinfo {pages} {030402} (\bibinfo {year}
  {2005})}\BibitemShut {NoStop}%
\bibitem [{\citenamefont {Smith}\ \emph {et~al.}(2006)\citenamefont {Smith},
  \citenamefont {Silberfarb}, \citenamefont {Deutsch},\ and\ \citenamefont
  {Jessen}}]{Smith2006}%
  \BibitemOpen
  \bibfield  {author} {\bibinfo {author} {\bibfnamefont {G.~A.}\ \bibnamefont
  {Smith}}, \bibinfo {author} {\bibfnamefont {A.}~\bibnamefont {Silberfarb}},
  \bibinfo {author} {\bibfnamefont {I.~H.}\ \bibnamefont {Deutsch}}, \ and\
  \bibinfo {author} {\bibfnamefont {P.~S.}\ \bibnamefont {Jessen}},\ }\href
  {\doibase 10.1103/PhysRevLett.97.180403} {\bibfield  {journal} {\bibinfo
  {journal} {Phys. Rev. Lett.}\ }\textbf {\bibinfo {volume} {97}},\ \bibinfo
  {pages} {180403} (\bibinfo {year} {2006})}\BibitemShut {NoStop}%
\bibitem [{\citenamefont {Riofrío}\ \emph {et~al.}(2011)\citenamefont
  {Riofrío}, \citenamefont {Jessen},\ and\ \citenamefont
  {Deutsch}}]{RiofrioJoPB2011}%
  \BibitemOpen
  \bibfield  {author} {\bibinfo {author} {\bibfnamefont {C.~A.}\ \bibnamefont
  {Riofrío}}, \bibinfo {author} {\bibfnamefont {P.~S.}\ \bibnamefont
  {Jessen}}, \ and\ \bibinfo {author} {\bibfnamefont {I.~H.}\ \bibnamefont
  {Deutsch}},\ }\href {http://stacks.iop.org/0953-4075/44/i=15/a=154007}
  {\bibfield  {journal} {\bibinfo  {journal} {Journal of Physics B: Atomic,
  Molecular and Optical Physics}\ }\textbf {\bibinfo {volume} {44}},\ \bibinfo
  {pages} {154007} (\bibinfo {year} {2011})}\BibitemShut {NoStop}%
\bibitem [{\citenamefont {Six}\ \emph {et~al.}(2016)\citenamefont {Six},
  \citenamefont {Campagne-Ibarcq}, \citenamefont {Dotsenko}, \citenamefont
  {Sarlette}, \citenamefont {Huard},\ and\ \citenamefont {Rouchon}}]{Six2016}%
  \BibitemOpen
  \bibfield  {author} {\bibinfo {author} {\bibfnamefont {P.}~\bibnamefont
  {Six}}, \bibinfo {author} {\bibfnamefont {P.}~\bibnamefont
  {Campagne-Ibarcq}}, \bibinfo {author} {\bibfnamefont {I.}~\bibnamefont
  {Dotsenko}}, \bibinfo {author} {\bibfnamefont {A.}~\bibnamefont {Sarlette}},
  \bibinfo {author} {\bibfnamefont {B.}~\bibnamefont {Huard}}, \ and\ \bibinfo
  {author} {\bibfnamefont {P.}~\bibnamefont {Rouchon}},\ }\href {\doibase
  10.1103/PhysRevA.93.012109} {\bibfield  {journal} {\bibinfo  {journal} {Phys.
  Rev. A}\ }\textbf {\bibinfo {volume} {93}},\ \bibinfo {pages} {012109}
  (\bibinfo {year} {2016})}\BibitemShut {NoStop}%
\bibitem [{\citenamefont {Shojaee}\ \emph {et~al.}(2018)\citenamefont
  {Shojaee}, \citenamefont {Jackson}, \citenamefont {Riofrio}, \citenamefont
  {Kalev},\ and\ \citenamefont {Deutsch}}]{Shojaee2018}%
  \BibitemOpen
  \bibfield  {author} {\bibinfo {author} {\bibfnamefont {E.}~\bibnamefont
  {Shojaee}}, \bibinfo {author} {\bibfnamefont {C.~S.}\ \bibnamefont
  {Jackson}}, \bibinfo {author} {\bibfnamefont {C.~A.}\ \bibnamefont
  {Riofrio}}, \bibinfo {author} {\bibfnamefont {A.}~\bibnamefont {Kalev}}, \
  and\ \bibinfo {author} {\bibfnamefont {I.~H.}\ \bibnamefont {Deutsch}},\
  }\href@noop {} {\bibfield  {journal} {\bibinfo  {journal} {arXiv:1805.01012}\
  } (\bibinfo {year} {2018})}\BibitemShut {NoStop}%
\bibitem [{\citenamefont {Blume-Kohout}(2010)}]{Blume-Kohout2010}%
  \BibitemOpen
  \bibfield  {author} {\bibinfo {author} {\bibfnamefont {R.}~\bibnamefont
  {Blume-Kohout}},\ }\href@noop {} {\bibfield  {journal} {\bibinfo  {journal}
  {New J. Phys.}\ }\textbf {\bibinfo {volume} {12}},\ \bibinfo {pages} {043034}
  (\bibinfo {year} {2010})}\BibitemShut {NoStop}%
\bibitem [{\citenamefont {Hradil}\ \emph {et~al.}(2004)\citenamefont {Hradil},
  \citenamefont {{\v R}eh{\'a}{\v c}ek}, \citenamefont {Fiur{\'a}{\v s}ek},\
  and\ \citenamefont {Je{\v z}ek}}]{Hradil2004}%
  \BibitemOpen
  \bibfield  {author} {\bibinfo {author} {\bibfnamefont {Z.}~\bibnamefont
  {Hradil}}, \bibinfo {author} {\bibfnamefont {J.}~\bibnamefont {{\v
  R}eh{\'a}{\v c}ek}}, \bibinfo {author} {\bibfnamefont {J.}~\bibnamefont
  {Fiur{\'a}{\v s}ek}}, \ and\ \bibinfo {author} {\bibfnamefont
  {M.}~\bibnamefont {Je{\v z}ek}},\ }in\ \href@noop {} {\emph {\bibinfo
  {booktitle} {Quantum {{State Estimation}}}}}\ (\bibinfo  {publisher}
  {{Springer, Berlin, Heidelberg}},\ \bibinfo {year} {2004})\ pp.\ \bibinfo
  {pages} {59--112}\BibitemShut {NoStop}%
\bibitem [{\citenamefont {Murch}\ \emph {et~al.}(2013)\citenamefont {Murch},
  \citenamefont {Weber}, \citenamefont {Macklin},\ and\ \citenamefont
  {Siddiqi}}]{Kater2013}%
  \BibitemOpen
  \bibfield  {author} {\bibinfo {author} {\bibfnamefont {K.~W.}\ \bibnamefont
  {Murch}}, \bibinfo {author} {\bibfnamefont {S.~J.}\ \bibnamefont {Weber}},
  \bibinfo {author} {\bibfnamefont {C.}~\bibnamefont {Macklin}}, \ and\
  \bibinfo {author} {\bibfnamefont {I.}~\bibnamefont {Siddiqi}},\ }\href@noop
  {} {\bibfield  {journal} {\bibinfo  {journal} {Nature}\ }\textbf {\bibinfo
  {volume} {502}},\ \bibinfo {pages} {211} (\bibinfo {year}
  {2013})}\BibitemShut {NoStop}%
\bibitem [{\citenamefont {Roch}\ \emph {et~al.}(2014)\citenamefont {Roch},
  \citenamefont {Schwartz}, \citenamefont {Motzoi}, \citenamefont {Macklin},
  \citenamefont {Vijay}, \citenamefont {Eddins}, \citenamefont {Korotkov},
  \citenamefont {Whaley}, \citenamefont {Sarovar},\ and\ \citenamefont
  {Siddiqi}}]{Roch2014}%
  \BibitemOpen
  \bibfield  {author} {\bibinfo {author} {\bibfnamefont {N.}~\bibnamefont
  {Roch}}, \bibinfo {author} {\bibfnamefont {M.~E.}\ \bibnamefont {Schwartz}},
  \bibinfo {author} {\bibfnamefont {F.}~\bibnamefont {Motzoi}}, \bibinfo
  {author} {\bibfnamefont {C.}~\bibnamefont {Macklin}}, \bibinfo {author}
  {\bibfnamefont {R.}~\bibnamefont {Vijay}}, \bibinfo {author} {\bibfnamefont
  {A.~W.}\ \bibnamefont {Eddins}}, \bibinfo {author} {\bibfnamefont {A.~N.}\
  \bibnamefont {Korotkov}}, \bibinfo {author} {\bibfnamefont {K.~B.}\
  \bibnamefont {Whaley}}, \bibinfo {author} {\bibfnamefont {M.}~\bibnamefont
  {Sarovar}}, \ and\ \bibinfo {author} {\bibfnamefont {I.}~\bibnamefont
  {Siddiqi}},\ }\href@noop {} {\bibfield  {journal} {\bibinfo  {journal} {Phys.
  Rev. Lett.}\ }\textbf {\bibinfo {volume} {112}},\ \bibinfo {pages} {170501}
  (\bibinfo {year} {2014})}\BibitemShut {NoStop}%
\bibitem [{\citenamefont {Campagne-Ibarcq}\ \emph {et~al.}(2016)\citenamefont
  {Campagne-Ibarcq}, \citenamefont {Six}, \citenamefont {Bretheau},
  \citenamefont {Sarlette}, \citenamefont {Mirrahimi}, \citenamefont
  {Rouchon},\ and\ \citenamefont {Huard}}]{ibarcq2015}%
  \BibitemOpen
  \bibfield  {author} {\bibinfo {author} {\bibfnamefont {P.}~\bibnamefont
  {Campagne-Ibarcq}}, \bibinfo {author} {\bibfnamefont {P.}~\bibnamefont
  {Six}}, \bibinfo {author} {\bibfnamefont {L.}~\bibnamefont {Bretheau}},
  \bibinfo {author} {\bibfnamefont {A.}~\bibnamefont {Sarlette}}, \bibinfo
  {author} {\bibfnamefont {M.}~\bibnamefont {Mirrahimi}}, \bibinfo {author}
  {\bibfnamefont {P.}~\bibnamefont {Rouchon}}, \ and\ \bibinfo {author}
  {\bibfnamefont {B.}~\bibnamefont {Huard}},\ }\href@noop {} {\bibfield
  {journal} {\bibinfo  {journal} {Phys. Rev. X}\ }\textbf {\bibinfo {volume}
  {6}},\ \bibinfo {pages} {011002} (\bibinfo {year} {2016})}\BibitemShut
  {NoStop}%
\bibitem [{\citenamefont {Weber}\ \emph {et~al.}(2016)\citenamefont {Weber},
  \citenamefont {Murch}, \citenamefont {Kimchi-Schwartz}, \citenamefont
  {Roch},\ and\ \citenamefont {Siddiqi}}]{Weberreview2016}%
  \BibitemOpen
  \bibfield  {author} {\bibinfo {author} {\bibfnamefont {S.~J.}\ \bibnamefont
  {Weber}}, \bibinfo {author} {\bibfnamefont {K.~W.}\ \bibnamefont {Murch}},
  \bibinfo {author} {\bibfnamefont {M.~E.}\ \bibnamefont {Kimchi-Schwartz}},
  \bibinfo {author} {\bibfnamefont {N.}~\bibnamefont {Roch}}, \ and\ \bibinfo
  {author} {\bibfnamefont {I.}~\bibnamefont {Siddiqi}},\ }\href {\doibase
  https://doi.org/10.1016/j.crhy.2016.07.007} {\bibfield  {journal} {\bibinfo
  {journal} {Comptes Rendus Physique}\ }\textbf {\bibinfo {volume} {17}},\
  \bibinfo {pages} {766 } (\bibinfo {year} {2016})},\ \bibinfo {note} {quantum
  microwaves / Micro-ondes quantiques}\BibitemShut {NoStop}%
\bibitem [{\citenamefont {Hacohen-Gourgy}\ \emph {et~al.}(2016)\citenamefont
  {Hacohen-Gourgy}, \citenamefont {Martin}, \citenamefont {Flurin},
  \citenamefont {Ramasesh}, \citenamefont {Whaley},\ and\ \citenamefont
  {Siddiqi}}]{Shay2016noncom}%
  \BibitemOpen
  \bibfield  {author} {\bibinfo {author} {\bibfnamefont {S.}~\bibnamefont
  {Hacohen-Gourgy}}, \bibinfo {author} {\bibfnamefont {L.~S.}\ \bibnamefont
  {Martin}}, \bibinfo {author} {\bibfnamefont {E.}~\bibnamefont {Flurin}},
  \bibinfo {author} {\bibfnamefont {V.~V.}\ \bibnamefont {Ramasesh}}, \bibinfo
  {author} {\bibfnamefont {K.~B.}\ \bibnamefont {Whaley}}, \ and\ \bibinfo
  {author} {\bibfnamefont {I.}~\bibnamefont {Siddiqi}},\ }\href@noop {}
  {\bibfield  {journal} {\bibinfo  {journal} {Nature}\ }\textbf {\bibinfo
  {volume} {538}},\ \bibinfo {pages} {491} (\bibinfo {year}
  {2016})}\BibitemShut {NoStop}%
\bibitem [{\citenamefont {Naghiloo}\ \emph {et~al.}(2016)\citenamefont
  {Naghiloo}, \citenamefont {Foroozani}, \citenamefont {Tan}, \citenamefont
  {Jadbabaie},\ and\ \citenamefont {Murch}}]{Naghiloo2016}%
  \BibitemOpen
  \bibfield  {author} {\bibinfo {author} {\bibfnamefont {M.}~\bibnamefont
  {Naghiloo}}, \bibinfo {author} {\bibfnamefont {N.}~\bibnamefont {Foroozani}},
  \bibinfo {author} {\bibfnamefont {D.}~\bibnamefont {Tan}}, \bibinfo {author}
  {\bibfnamefont {A.}~\bibnamefont {Jadbabaie}}, \ and\ \bibinfo {author}
  {\bibfnamefont {K.~W.}\ \bibnamefont {Murch}},\ }\href@noop {} {\bibfield
  {journal} {\bibinfo  {journal} {Nature Communications}\ }\textbf {\bibinfo
  {volume} {7}},\ \bibinfo {pages} {11527} (\bibinfo {year}
  {2016})}\BibitemShut {NoStop}%
\bibitem [{\citenamefont {Barchielli}\ and\ \citenamefont
  {Gregoratti}(2009)}]{BookBachielli}%
  \BibitemOpen
  \bibfield  {author} {\bibinfo {author} {\bibfnamefont {A.}~\bibnamefont
  {Barchielli}}\ and\ \bibinfo {author} {\bibfnamefont {M.}~\bibnamefont
  {Gregoratti}},\ }\href@noop {} {\emph {\bibinfo {title} {Quantum trajectories
  and measurements in continuous time}}}\ (\bibinfo  {publisher}
  {Springer-Verlag Berlin Heidelberg},\ \bibinfo {year} {2009})\BibitemShut
  {NoStop}%
\bibitem [{\citenamefont {Carmichael}(1993)}]{BookCarmichael}%
  \BibitemOpen
  \bibfield  {author} {\bibinfo {author} {\bibfnamefont {H.~J.}\ \bibnamefont
  {Carmichael}},\ }\href@noop {} {\emph {\bibinfo {title} {An Open Systems
  Approach to Quantum Optics}}}\ (\bibinfo  {publisher} {Springer, Berlin},\
  \bibinfo {year} {1993})\BibitemShut {NoStop}%
\bibitem [{\citenamefont {Wiseman}\ and\ \citenamefont
  {Milburn}(2009)}]{wiseman2009quantum}%
  \BibitemOpen
  \bibfield  {author} {\bibinfo {author} {\bibfnamefont {H.~M.}\ \bibnamefont
  {Wiseman}}\ and\ \bibinfo {author} {\bibfnamefont {G.~J.}\ \bibnamefont
  {Milburn}},\ }\href@noop {} {\emph {\bibinfo {title} {Quantum measurement and
  control}}}\ (\bibinfo  {publisher} {Cambridge University Press},\ \bibinfo
  {year} {2009})\BibitemShut {NoStop}%
\bibitem [{\citenamefont {Jacobs}(2014)}]{BookJacobs}%
  \BibitemOpen
  \bibfield  {author} {\bibinfo {author} {\bibfnamefont {K.}~\bibnamefont
  {Jacobs}},\ }\href@noop {} {\emph {\bibinfo {title} {Quantum Measurement
  Theory and its Applications}}}\ (\bibinfo  {publisher} {Cambridge University
  Press},\ \bibinfo {year} {2014})\BibitemShut {NoStop}%
\bibitem [{\citenamefont {Chantasri}\ \emph {et~al.}(2013)\citenamefont
  {Chantasri}, \citenamefont {Dressel},\ and\ \citenamefont
  {Jordan}}]{Chantasri2013}%
  \BibitemOpen
  \bibfield  {author} {\bibinfo {author} {\bibfnamefont {A.}~\bibnamefont
  {Chantasri}}, \bibinfo {author} {\bibfnamefont {J.}~\bibnamefont {Dressel}},
  \ and\ \bibinfo {author} {\bibfnamefont {A.~N.}\ \bibnamefont {Jordan}},\
  }\href@noop {} {\bibfield  {journal} {\bibinfo  {journal} {Phys. Rev. A}\
  }\textbf {\bibinfo {volume} {88}},\ \bibinfo {pages} {042110} (\bibinfo
  {year} {2013})}\BibitemShut {NoStop}%
\bibitem [{\citenamefont {Chantasri}\ and\ \citenamefont
  {Jordan}(2015)}]{chantasri2015stochastic}%
  \BibitemOpen
  \bibfield  {author} {\bibinfo {author} {\bibfnamefont {A.}~\bibnamefont
  {Chantasri}}\ and\ \bibinfo {author} {\bibfnamefont {A.~N.}\ \bibnamefont
  {Jordan}},\ }\href@noop {} {\bibfield  {journal} {\bibinfo  {journal}
  {Physical Review A}\ }\textbf {\bibinfo {volume} {92}},\ \bibinfo {pages}
  {032125} (\bibinfo {year} {2015})}\BibitemShut {NoStop}%
\bibitem [{\citenamefont {Gross}\ \emph {et~al.}(2015)\citenamefont {Gross},
  \citenamefont {Dangniam}, \citenamefont {Ferrie},\ and\ \citenamefont
  {Caves}}]{Gross2015}%
  \BibitemOpen
  \bibfield  {author} {\bibinfo {author} {\bibfnamefont {J.~A.}\ \bibnamefont
  {Gross}}, \bibinfo {author} {\bibfnamefont {N.}~\bibnamefont {Dangniam}},
  \bibinfo {author} {\bibfnamefont {C.}~\bibnamefont {Ferrie}}, \ and\ \bibinfo
  {author} {\bibfnamefont {C.~M.}\ \bibnamefont {Caves}},\ }\href {\doibase
  10.1103/PhysRevA.92.062133} {\bibfield  {journal} {\bibinfo  {journal} {Phys.
  Rev. A}\ }\textbf {\bibinfo {volume} {92}},\ \bibinfo {pages} {062133}
  (\bibinfo {year} {2015})}\BibitemShut {NoStop}%
\bibitem [{\citenamefont {Wiseman}(1994)}]{Wiseman1994}%
  \BibitemOpen
  \bibfield  {author} {\bibinfo {author} {\bibfnamefont {H.~M.}\ \bibnamefont
  {Wiseman}},\ }\href@noop {} {\bibfield  {journal} {\bibinfo  {journal} {Phys.
  Rev. A}\ }\textbf {\bibinfo {volume} {49}},\ \bibinfo {pages} {2133}
  (\bibinfo {year} {1994})}\BibitemShut {NoStop}%
\bibitem [{\citenamefont {Doherty}\ \emph {et~al.}(2000)\citenamefont
  {Doherty}, \citenamefont {Habib}, \citenamefont {Jacobs}, \citenamefont
  {Mabuchi},\ and\ \citenamefont {Tan}}]{Doherty2000}%
  \BibitemOpen
  \bibfield  {author} {\bibinfo {author} {\bibfnamefont {A.~C.}\ \bibnamefont
  {Doherty}}, \bibinfo {author} {\bibfnamefont {S.}~\bibnamefont {Habib}},
  \bibinfo {author} {\bibfnamefont {K.}~\bibnamefont {Jacobs}}, \bibinfo
  {author} {\bibfnamefont {H.}~\bibnamefont {Mabuchi}}, \ and\ \bibinfo
  {author} {\bibfnamefont {S.~M.}\ \bibnamefont {Tan}},\ }\href@noop {}
  {\bibfield  {journal} {\bibinfo  {journal} {Phys. Rev. A}\ }\textbf {\bibinfo
  {volume} {62}},\ \bibinfo {pages} {012105} (\bibinfo {year}
  {2000})}\BibitemShut {NoStop}%
\bibitem [{\citenamefont {Wheatley}\ \emph {et~al.}(2010)\citenamefont
  {Wheatley}, \citenamefont {Berry}, \citenamefont {Yonezawa}, \citenamefont
  {Nakane}, \citenamefont {Arao}, \citenamefont {Pope}, \citenamefont {Ralph},
  \citenamefont {Wiseman}, \citenamefont {Furusawa},\ and\ \citenamefont
  {Huntington}}]{whea10}%
  \BibitemOpen
  \bibfield  {author} {\bibinfo {author} {\bibfnamefont {T.~A.}\ \bibnamefont
  {Wheatley}}, \bibinfo {author} {\bibfnamefont {D.~W.}\ \bibnamefont {Berry}},
  \bibinfo {author} {\bibfnamefont {H.}~\bibnamefont {Yonezawa}}, \bibinfo
  {author} {\bibfnamefont {D.}~\bibnamefont {Nakane}}, \bibinfo {author}
  {\bibfnamefont {H.}~\bibnamefont {Arao}}, \bibinfo {author} {\bibfnamefont
  {D.~T.}\ \bibnamefont {Pope}}, \bibinfo {author} {\bibfnamefont {T.~C.}\
  \bibnamefont {Ralph}}, \bibinfo {author} {\bibfnamefont {H.~M.}\ \bibnamefont
  {Wiseman}}, \bibinfo {author} {\bibfnamefont {A.}~\bibnamefont {Furusawa}}, \
  and\ \bibinfo {author} {\bibfnamefont {E.~H.}\ \bibnamefont {Huntington}},\
  }\href {\doibase 10.1103/PhysRevLett.104.093601} {\bibfield  {journal}
  {\bibinfo  {journal} {Phys. Rev. Lett.}\ }\textbf {\bibinfo {volume} {104}},\
  \bibinfo {pages} {093601} (\bibinfo {year} {2010})}\BibitemShut {NoStop}%
\bibitem [{\citenamefont {Yonezawa}\ \emph {et~al.}(2012)\citenamefont
  {Yonezawa}, \citenamefont {Nakane}, \citenamefont {Wheatley}, \citenamefont
  {Iwasawa}, \citenamefont {Takeda}, \citenamefont {Arao}, \citenamefont
  {Ohki}, \citenamefont {Tsumura}, \citenamefont {Berry}, \citenamefont
  {Ralph}, \citenamefont {Wiseman}, \citenamefont {Huntington},\ and\
  \citenamefont {Furusawa}}]{Yonezawa1514}%
  \BibitemOpen
  \bibfield  {author} {\bibinfo {author} {\bibfnamefont {H.}~\bibnamefont
  {Yonezawa}}, \bibinfo {author} {\bibfnamefont {D.}~\bibnamefont {Nakane}},
  \bibinfo {author} {\bibfnamefont {T.~A.}\ \bibnamefont {Wheatley}}, \bibinfo
  {author} {\bibfnamefont {K.}~\bibnamefont {Iwasawa}}, \bibinfo {author}
  {\bibfnamefont {S.}~\bibnamefont {Takeda}}, \bibinfo {author} {\bibfnamefont
  {H.}~\bibnamefont {Arao}}, \bibinfo {author} {\bibfnamefont {K.}~\bibnamefont
  {Ohki}}, \bibinfo {author} {\bibfnamefont {K.}~\bibnamefont {Tsumura}},
  \bibinfo {author} {\bibfnamefont {D.~W.}\ \bibnamefont {Berry}}, \bibinfo
  {author} {\bibfnamefont {T.~C.}\ \bibnamefont {Ralph}}, \bibinfo {author}
  {\bibfnamefont {H.~M.}\ \bibnamefont {Wiseman}}, \bibinfo {author}
  {\bibfnamefont {E.~H.}\ \bibnamefont {Huntington}}, \ and\ \bibinfo {author}
  {\bibfnamefont {A.}~\bibnamefont {Furusawa}},\ }\href {\doibase
  10.1126/science.1225258} {\bibfield  {journal} {\bibinfo  {journal}
  {Science}\ }\textbf {\bibinfo {volume} {337}},\ \bibinfo {pages} {1514}
  (\bibinfo {year} {2012})}\BibitemShut {NoStop}%
\bibitem [{\citenamefont {Ralph}\ \emph {et~al.}(2011)\citenamefont {Ralph},
  \citenamefont {Jacobs},\ and\ \citenamefont {Hill}}]{ralph2011frequency}%
  \BibitemOpen
  \bibfield  {author} {\bibinfo {author} {\bibfnamefont {J.~F.}\ \bibnamefont
  {Ralph}}, \bibinfo {author} {\bibfnamefont {K.}~\bibnamefont {Jacobs}}, \
  and\ \bibinfo {author} {\bibfnamefont {C.~D.}\ \bibnamefont {Hill}},\
  }\href@noop {} {\bibfield  {journal} {\bibinfo  {journal} {Physical Review
  A}\ }\textbf {\bibinfo {volume} {84}},\ \bibinfo {pages} {052119} (\bibinfo
  {year} {2011})}\BibitemShut {NoStop}%
\bibitem [{\citenamefont {Kiilerich}\ and\ \citenamefont
  {M\o{}lmer}(2016)}]{Kiilerich2016}%
  \BibitemOpen
  \bibfield  {author} {\bibinfo {author} {\bibfnamefont {A.~H.}\ \bibnamefont
  {Kiilerich}}\ and\ \bibinfo {author} {\bibfnamefont {K.}~\bibnamefont
  {M\o{}lmer}},\ }\href {\doibase 10.1103/PhysRevA.94.032103} {\bibfield
  {journal} {\bibinfo  {journal} {Phys. Rev. A}\ }\textbf {\bibinfo {volume}
  {94}},\ \bibinfo {pages} {032103} (\bibinfo {year} {2016})}\BibitemShut
  {NoStop}%
\bibitem [{\citenamefont {Cortez}\ \emph {et~al.}(2017)\citenamefont {Cortez},
  \citenamefont {Chantasri}, \citenamefont {Garc\'{\i}a-Pintos}, \citenamefont
  {Dressel},\ and\ \citenamefont {Jordan}}]{Luisparaest2017}%
  \BibitemOpen
  \bibfield  {author} {\bibinfo {author} {\bibfnamefont {L.}~\bibnamefont
  {Cortez}}, \bibinfo {author} {\bibfnamefont {A.}~\bibnamefont {Chantasri}},
  \bibinfo {author} {\bibfnamefont {L.~P.}\ \bibnamefont {Garc\'{\i}a-Pintos}},
  \bibinfo {author} {\bibfnamefont {J.}~\bibnamefont {Dressel}}, \ and\
  \bibinfo {author} {\bibfnamefont {A.~N.}\ \bibnamefont {Jordan}},\ }\href
  {\doibase 10.1103/PhysRevA.95.012314} {\bibfield  {journal} {\bibinfo
  {journal} {Phys. Rev. A}\ }\textbf {\bibinfo {volume} {95}},\ \bibinfo
  {pages} {012314} (\bibinfo {year} {2017})}\BibitemShut {NoStop}%
\bibitem [{\citenamefont {Ralph}\ \emph {et~al.}(2017)\citenamefont {Ralph},
  \citenamefont {Maskell},\ and\ \citenamefont {Jacobs}}]{Ralph2017}%
  \BibitemOpen
  \bibfield  {author} {\bibinfo {author} {\bibfnamefont {J.~F.}\ \bibnamefont
  {Ralph}}, \bibinfo {author} {\bibfnamefont {S.}~\bibnamefont {Maskell}}, \
  and\ \bibinfo {author} {\bibfnamefont {K.}~\bibnamefont {Jacobs}},\ }\href
  {\doibase 10.1103/PhysRevA.96.052306} {\bibfield  {journal} {\bibinfo
  {journal} {Phys. Rev. A}\ }\textbf {\bibinfo {volume} {96}},\ \bibinfo
  {pages} {052306} (\bibinfo {year} {2017})}\BibitemShut {NoStop}%
\bibitem [{\citenamefont {Di\'osi}(1988)}]{Diosi1988}%
  \BibitemOpen
  \bibfield  {author} {\bibinfo {author} {\bibfnamefont {L.}~\bibnamefont
  {Di\'osi}},\ }\href@noop {} {\bibfield  {journal} {\bibinfo  {journal} {Phys.
  Lett. A}\ }\textbf {\bibinfo {volume} {129}},\ \bibinfo {pages} {419}
  (\bibinfo {year} {1988})}\BibitemShut {NoStop}%
\bibitem [{\citenamefont {Lindblad}(1976)}]{Lind1976}%
  \BibitemOpen
  \bibfield  {author} {\bibinfo {author} {\bibfnamefont {G.}~\bibnamefont
  {Lindblad}},\ }\href@noop {} {\bibfield  {journal} {\bibinfo  {journal}
  {Commun. Math. Phys.}\ }\textbf {\bibinfo {volume} {48}},\ \bibinfo {pages}
  {119} (\bibinfo {year} {1976})}\BibitemShut {NoStop}%
\bibitem [{\citenamefont {Korotkov}(1999)}]{Korotkov1999}%
  \BibitemOpen
  \bibfield  {author} {\bibinfo {author} {\bibfnamefont {A.~N.}\ \bibnamefont
  {Korotkov}},\ }\href@noop {} {\bibfield  {journal} {\bibinfo  {journal}
  {Phys. Rev. B}\ }\textbf {\bibinfo {volume} {60}},\ \bibinfo {pages} {5737}
  (\bibinfo {year} {1999})}\BibitemShut {NoStop}%
\bibitem [{\citenamefont {Korotkov}(2001)}]{Korotkov2001}%
  \BibitemOpen
  \bibfield  {author} {\bibinfo {author} {\bibfnamefont {A.~N.}\ \bibnamefont
  {Korotkov}},\ }\href@noop {} {\bibfield  {journal} {\bibinfo  {journal}
  {Phys. Rev. B}\ }\textbf {\bibinfo {volume} {63}},\ \bibinfo {pages} {115403}
  (\bibinfo {year} {2001})}\BibitemShut {NoStop}%
\bibitem [{\citenamefont {Cram{\'e}r}(1946)}]{Cramer1946}%
  \BibitemOpen
  \bibfield  {author} {\bibinfo {author} {\bibfnamefont {H.}~\bibnamefont
  {Cram{\'e}r}},\ }\href@noop {} {\emph {\bibinfo {title} {Mathematical
  {{Methods}} of {{Statistics}}}}}\ (\bibinfo  {publisher} {{Princeton
  University Press}},\ \bibinfo {address} {Princeton},\ \bibinfo {year}
  {1946})\BibitemShut {NoStop}%
\bibitem [{\citenamefont {Gamel}(2016)}]{Omar2016}%
  \BibitemOpen
  \bibfield  {author} {\bibinfo {author} {\bibfnamefont {O.}~\bibnamefont
  {Gamel}},\ }\href {\doibase 10.1103/PhysRevA.93.062320} {\bibfield  {journal}
  {\bibinfo  {journal} {Phys. Rev. A}\ }\textbf {\bibinfo {volume} {93}},\
  \bibinfo {pages} {062320} (\bibinfo {year} {2016})}\BibitemShut {NoStop}%
\end{thebibliography}
%



\appendix

\section{Analytic calculation for the Fisher information matrix}\label{sec-fisher}

In this section, we present a detailed calculation of the Fisher information matrix Eq.~\eqref{eq-Fishermat2} for estimating a two-qubit unknown state in the limit of weak measurement. As defined in the main text, the weak measurement
operator for a short time $\delt$ can be written as
\begin{equation}
\begin{aligned}\m= \,& \text{\ensuremath{\Big(\frac{\delt}{2\pi\var}\Big)}}^{\frac{1}{4}}\exp\left[-\frac{(\rk-\sz)^{2}\delt}{4\var}\right],\\
= \,& \e^{-\frac{\delt}{4\var}}\sqrt{\fr}\exp\Big(\frac{\rk\delt}{2\var}\sz\Big),
\end{aligned}
\end{equation}
where $\sz = \ZI$ and
\begin{equation}
\fr=\sqrt{\frac{\delt}{2\pi\var}}\exp\Big(-\frac{\rks}{2\var}\delt\Big),\label{eq:fr}
\end{equation}
is a Gaussian distribution of $\rk$ with mean zero and variance $\var/\delt$.
The trajectory Kraus operator is
\begin{equation}
\ot=\m[n]\uk\cdots\m[1]\uk,\label{eq:kraus op}
\end{equation}
where we denote the total time of the measurement as $T=n\,\delt$.

When $\delt$ is very small, we can expand each $\m$ to the first
order of $\delt$,
\begin{equation}
\m\approx\e^{-\frac{\delt}{4\var}}\sqrt{\fr}\Big(\ib+\frac{\rk\delt}{2\var}\sz\Big),\label{eq:appr meas op}
\end{equation}
where $\ib=\id$. Note that we only
 need the first order of $\delt$ in Eq.~(\ref{eq:appr meas op}),
even though we want to keep higher order terms in $1/\tau$ in $\ot$ later. This is because higher order terms in $\delt$ in (\ref{eq:appr meas op}) will have measure zero when we perform integrations over time (see the following equations). The time integration needs a limit taking $\delt\rightarrow0$ and $n\rightarrow\infty$ for any finite $T$. We substitute the measurement operators and the unitary operators in Eq.~\eqref{eq:kraus op}, keeping only to first order in $\dt$. We then replace any time summations $\sum_{k}\delt$ with the integral $\int\dtp$ to obtain the operator $\ot$ in a time-ordering integral form:
\begin{equation}
\begin{aligned}\otr= & \e^{-\frac{T}{4\var}}\sqrt{\prod_{k}\frtp[t_k]}\Big[\uk[T]+\sum_{k=1}^{+\infty}\frac{1}{(2\var)^{k}k!}\\
 & \mt\mint\rtp[t_1]\cdots\rtp[t_k]\uk[t-{{\tp[k]}}]\sz\uk[{{\tp[k]}}-{{\tp[k-1]}}]\\
 & \sz\cdots\uk[{{\tp[2]}}-{{\tp[1]}}]\sz\uk[{\tp[1]}]\dtp[1]\cdots\dtp[k]\Big],
\end{aligned}
\end{equation}
where $\mt$ is the time-ordering operator and $\rtp[t_k]$ is a measurement result at time $\tp[k]$.
If we denote a time-dependent operator
\begin{equation}
\zt=\ukd[t]\sz\uk[t],
\end{equation}
then $\otr$ can be simplified to
\begin{equation}
\begin{aligned}\otr=&  \e^{-\frac{T}{4\var}}\sqrt{\prod_{k}\frtp[t_k]}\uk[T]\Big[\ib+\sum_{k=1}^{+\infty}\frac{1}{(2\var)^{k}k!}\\
 &\times \mt\mint\rtp[t_1]\cdots\rtp[t_k]\ztp[k]\cdots\ztp[1]\dtp[1]\cdots\dtp[k]\Big],
\end{aligned}
\end{equation}
and we can write a POVM element corresponding to $\otr$ as
\begin{equation}
\etr=\otr[\dagger]\otr.
\end{equation}

In the weak measurement limit, we will keep the terms up to the order $1/\var$ throughout
the rest of this appendix. For the POVM $\etr$, we expand to first order in $1/\var$ and get,
\begin{equation}
\etr=\e^{-\frac{T}{2\var}}\prod_{k}\frtp[t_k]\left[\ib+\frac{1}{\var}\sint\rtp[t]\ztp\dtp\right],
\end{equation}
where $\rtp[t]$ is the measurement result at time $\tp$. Given an initial state of the two qubits
\begin{equation}
\ri=\frac{1}{4}\sum_{i,j=0,x,y,z}\cij\sij,
\end{equation}
where ${\hat \sigma}_{0}={\hat I}$ and $\cij[00]=1$, the probability for
the measurement record $\rt$ can be computed from the POVM element,
\begin{equation}
\begin{aligned}\pr= & \tr(\etr\ri),\\
= & \e^{-\frac{T}{2\var}}\prod_k \frtp[t_k]\left[1+\frac{1}{\var}\sint\rtp[t]\avgo{{\ztp}}\dtp\right],
\end{aligned}
\end{equation}
where $\avgo{\cdot}$ denotes the expectation value with the initial
state $\ri$. We calculate the derivative of $\pr$ with respect to the parameter
$\cij$ giving,
\begin{equation}
\pc\pr=\frac{\e^{-\frac{T}{2\var}}}{4\tau}\prod_k\frtp[t_k]\sint\rtp[t]\aij\dtp,
\end{equation}
where we have defined,
\begin{equation}
\aij=\tr(\ztp\sij).
\end{equation}

We then use the above equations to compute the $(ij,\,\ijp)$-th element of Fisher information matrix for estimating a two-qubit unknown state,
\begin{equation}
\fij=\dotsint\frac{\pc\pr\pc[\ijp]\pr}{\pr}\drtp[t_1]\cdots\drtp[t_n].\label{eq:fish}
\end{equation}
Keeping to first order in $1/\var$, we have
\begin{equation}
\frac{1}{\pr}=\frac{\e^{\frac{T}{2\var}}}{\prod_k\frtp[t_k]}\left[1-\frac{1}{\var}\sint\rtp[t]\avgo{{\ztp}}\dtp\right],
\end{equation}
and
\begin{equation}
\begin{aligned} & \frac{\pc\pr\pc[\ijp]\pr}{\pr}\\
 & =\frac{\e^{-\frac{T}{2\var}}}{16\var[2]}\prod_k \frtp[t_k]\sint\sint\rtp[t_1]\rtp[t_2]\aij[][1]\aij[\ijp][2]\dtp[1]\dtp[2].
\end{aligned}
\label{eq:2}
\end{equation}
Substituting the above two equations to the Fisher information matrix element Eq.~\eqref{eq:fish}, we then perform the integrals of $r_1, r_2,..., r_n$ with $\frtp[k]$ being a Gaussian distribution with zero mean and variance $\var/\dt$. Using the properties of integrals with Gaussian functions, we have
\begin{equation}
\iint\prod_k\frtp[t_k]\rtp[t_1]\rtp[t_2]\drtp[t_1]\cdots\drtp[t_n]=\var\dett[1][2].\label{eq:int}
\end{equation}
By applying Eq. (\ref{eq:int}) to (\ref{eq:2}), we obtain
\begin{equation}
\begin{aligned} & \iint\prod_k\frtp[t_k]\rtp[t_1]\rtp[t_2]\aij[][1]\aij[\ijp][2]\drtp[t_1]\cdots\drtp[t_n]\\
 & =\var\aij[][1]\aij[\ijp][2]\dett[1][2],
\end{aligned}
\end{equation}
and, the $(ij,\,\ijp)$-th element of Fisher information matrix
\begin{equation}
\fij=\frac{\e^{-\frac{T}{2\var}}}{16\var}\int_{0}^{T}\aij\aij[\ijp]\dtp,
\end{equation}
where we have replaced $\tp[1]$ or $\tp[2]$ by $\tp$.

\section{Methods for numerical simulation}\label{sec-numer}

Numerical simulation is used to investigate the state estimation quality for different measurement and control settings. There are two main parts in the numerical calculation. The first part is to generate $N$ measurement records using an initial state $\rho_0$ chosen at random, regardless of its purity or levels of entanglement. Starting with the given initial state, each record consists of measurement results at all time steps $r_1, r_2, ..., r_n$ are randomly generated with probability distributions $P(r_k|\rho_k)$ for $k = 1, 2,..., n$, where the quantum states $\rho_k$ are calculated for each time step with an update equation,
\begin{align}
\rho(t_{k+1}) = \frac{U(\dt) M(r_k) \rho(t_k) M(r_k)^{\dagger}U(\dt)^{\dagger}}{{\rm Tr}[U(\dt)M(r_k) \rho(t_k) M(r_k)^{\dagger}U(\dt)^{\dagger}]},
\end{align}
where $M(r_k)$ is defined in Eq.~\eqref{eq-measop} in the main text and $U(\dt) = \exp(-i {\hat H} \dt)$. We use the time step $\dt = 0.01$ and the total measurement time is fixed at $T = 2$ for all results presented in the main text. The second part of the simulation is to use the generated $N$ records in an estimation of the initial state $\rho_0$ assuming that it is unknown. Using the $N$ records denoted as $R_1$, $R_2$,..., $R_N$, we construct evolution operators ${\cal M}_{R_j}$ for $j = 1,..., N$ defined in Eq.~\eqref{eq-stateupdate} and compute a function $P(\{ R \} | \rho')$ Eq.~\eqref{eq-likelihoodR} for an arbitrary state $\rho'$. This is the likelihood function used in estimating the unknown state.

For the single-qubit state tomography, we focus more on the Bayesian estimator because of its robustness in comparison to the maximum likelihood estimation. The probability function $P(\{ R\} | \rho')$ is then applied to $10^4$ different trial states $\rho'$ which are equally distributed on the Bloch sphere. We use a MATLAB code named ``RandomDensityMatrix'' from QETLAB to generate single-qubit random density matrices distributed uniformly according to Hilbert-Schmidt measure [http://www.qetlab.com]. The reason we chose $10^4$ trial states is so that an average distance between any two trial state is given as $\Delta \rho = \left( \frac{4}{3}\frac{\pi}{10000} \right)^{1/3} = 0.074 \le 0.1$. Once the probability function is calculated for every trial state, we find a Bayesian probability from,
\begin{align}
P(\rho' | \{ R\}) = P(\{ R \}|\rho') P(\rho'),
\end{align}
where, in the case of the uniform distribution, we simply replace $P(\rho')$ by 1. This Bayesian probability is used to calculate the Bayesian estimator Eq.~\eqref{eq-BME} and the maximum likelihood estimator Eq.~\eqref{eq-MLE}. The error of the estimation is computed from the Bayesian covariance matrix Eq.~\eqref{eq-baycovmat} using the true state $\rho_0$. For the results in Figure~\ref{fig-sqbRabi}(b) and (c), we generated 100 sets of the estimators, from different sets of $N$ records, so that the results are more reliable.

For the remote-qubit state tomography, we use the similar method as in the single qubit case. The trial states in this case are $4\times 4$ density matrices, which are tensor products of the ancilla initial state and the single-qubit trial state $\rho'$ (same $10^4$ set as in the single-qubit case). For the results shown in Figure~\ref{fig-rqbresults}, we compute the root mean square errors from 100 sets of the estimators, considering each element of the density matrix as an estimated parameter. We could use the errors from the Bayesian covariance matrix similar to the single qubit case, but using this type of errors makes the transition to the two-qubit tomography discussion easier to understand. The results for single-qubit and remote-qubit cases are generated using Python and MATLAB codes.

For the two-qubit state tomography, we use a different technique from the previous two examples because the state space of the two qubits is much larger. As stated in the main text, in order to use the Bayesian estimator for the two-qubit tomography, one needs much more than $10^8$ trial states. This number comes from a number of product states of the single-qubit trial states $10^4 \times 10^4$. Instead of using the Bayesian estimator, we then compute the maximum likelihood estimator from maximizing the function $P(\{ R \}|\rho')$ over all valid two-qubit states $\rho'$, as discussed in the main text. We use the built-in Differential Evolution algorithm in Mathematica to search for the estimator, using the constraints in Eq.~\eqref{eq-constraints} and 50 searches initialized with random numbers between $-\sqrt{3}$ and $\sqrt{3}$ (for each of the 15 elements, see the first constraint in Eq.~\eqref{eq-constraints}). We compute 50 MLE estimators, each with different sets of $N = 4000$ measurement records, and calculate the RMSEs from the 50 estimators to get the results in Figure~\ref{fig-tqbresults}.

\section{Lists of random initial states of two-qubit tomography simulation}\label{sec-randomint}

For the single qubit case, the average fidelity for different initial states in Figure~\ref{fig-sqbRabi}(c) are computed from these nine randomly chosen states: $\{x_0, y_0, z_0\}=$ \{-0.4, -0.6, 0.3\}, \{-0.4, 0.6, -0.3\}, \{-0.4, 0.6, 0.3\}, \{0.4, 0.6, -0.3\}, \{-0.7, -0.5 , -0.3\}, \{-0.7, -0.5 ,0.3\}, \{0.7, -0.5, 0.3\}, \{-0.5, 0.3, 0.8\}, \{-0.5, -0.3, -0.8\}, \{0.5, 0.3, 0.8\}.

For the remote qubit, the average fidelity for different initial states mentioned in the caption of Figure~\ref{fig-rqbresults}(d) are computed from these ten randomly chosen states: $\{x_{\rm u}^0, y_{\rm u}^0, z_{\rm u}^0\}$ = \{$1/\sqrt{2}$, $1/\sqrt{2}$, 0\}, \{-0.6, -0.4, 0.3\}, \{-0.7, -0.5, -0.3\}, \{0.0, -0.5, 0.3\}, \{0.3, -0.3, 0.3\}, \{0.3, -0.7, 0.5\}, \{0.5, 0.3, 0.8\}, \{0.7, -0.5, 0.1\}, \{0.7, 0.0, 0.0\}, \{0.7, 0.0, 0.3\}. The fidelities for these initial states (each with 100 repetitions) are 0.98853, 0.998968, 0.998952, 0.999116, 0.999447, 0.99888, 0.993246, 0.999316, 0.999317, 0.999122, giving an average $0.998 \pm 0.004$.

In the two-qubit case, we presented in the main text the average fidelities in Figure~\ref{fig-tqbresults}. The initial states were chosen randomly including mixed states and non-separable states. We define a function $W(p, \rho_1, \rho_2)$,
\begin{align}
W(p, \rho_1, \rho_2) = (1-p) \rho_1 \otimes \rho_2 + p |\Phi^+\ra \la \Phi^+|,
\end{align}
where $|\Phi^+\ra = 1/\sqrt{2} (|00\ra + |11\ra)$ is a Bell state written in a computational basis $00,01,10,11$ of the two-qubit state.
The nine random initial states are,
\begin{itemize}
\item $W(0.5, \frac{1}{2}(\I + 0.7\X -0.2 \Y + 0.3\Z), \frac{1}{2}(\I + 0.6\X -0.1 \Y + 0.4\Z))$
\item $|\Phi^+\ra \la \Phi^+|$
\item $W(0.2, \frac{1}{2}(\I - 0.4\X -0.75 \Y + 0.5\Z), \frac{1}{2}(\I + 0.6\X -0.5 \Y + 0.6\Z))$
\item $W(0.5, \frac{1}{2}(\I + 0.2\X -0.75 \Y + 0.5\Z), \frac{1}{2}(\I + 0.6\X -0.5 \Y + 0.4\Z))$
\item $W(0.8, \frac{1}{2}(\I + 0.2\X -0.75 \Y + 0.5\Z), \frac{1}{2}(\I + 0.6\X -0.5 \Y + 0.4\Z))$
\item $\frac{1}{2}(\I + 0.7\X -0.2 \Y + 0.5\Z) \otimes \frac{1}{2}(\I + 0.6\X -0.5 \Y + 0.4\Z)$
\item $W(0.5, \frac{1}{2}(\I + 0.7\X -0.2 \Y + 0.5\Z), \frac{1}{2}(\I + 0.6\X -0.5 \Y + 0.4\Z)$
\item $W(0.8, \frac{1}{2}(\I + 0.7\X -0.2 \Y + 0.5\Z), \frac{1}{2}(\I + 0.6\X -0.5 \Y + 0.4\Z)$
\item $\frac{1}{2}(\I + 0.7\X -0.2 \Y + 0.3\Z) \otimes \frac{1}{2}(\I + 0.6\X -0.1 \Y + 0.4\Z)$
\end{itemize}

\end{document}